\documentclass[12pt]{amsart}
\pdfoutput=1

\usepackage{hyperref}

\usepackage{amsmath}
\usepackage[norelsize]{algorithm2e}
\usepackage{scalerel}
\usepackage{tikz-cd}
\usepackage{ulem}
\usepackage{float}
\restylefloat{figure}
\usepackage{stackengine}
\usepackage[english]{babel}
\usepackage[light,condensed,math]{anttor}
\usepackage[T1]{fontenc}
\usepackage{kpfonts}
\usepackage{comment}
\usepackage{array}
\usepackage{lmodern}
\usepackage[mathscr]{euscript}
\usepackage[utf8x]{inputenc}
\usepackage{breqn}
\usepackage{todonotes}
\usepackage{slashed}
\usepackage{youngtab}
\usepackage{amsmath}
\usepackage{amssymb}
\usepackage{amsxtra}
\usepackage{color}
\usepackage[margin=1.2in]{geometry}

\newcommand {\3}{\underline{\bf 3}}
\newcommand {\4}{\underline{\bf 4}}
\newcommand {\6}{\underline{\bf 6}}
\usepackage{float}
\restylefloat{figure}
\newcommand{\boxit}[1]{\vbox{\hrule\hbox{\vrule\kern8pt
\vbox{\hbox{\kern8pt}\hbox{\vbox{#1}}\hbox{\kern8pt}}
\kern8pt\vrule}\hrule}}
\newcommand{\mathboxit}[1]{\vbox{\hrule\hbox{\vrule\kern8pt\vbox{\kern8pt
\hbox{$\displaystyle #1$}\kern8pt}\kern8pt\vrule}\hrule}}
\newcommand{\picti}[2]{\includegraphics[width=#1cm]{#2.pdf}}

\makeatletter

\DeclareFontFamily{OMX}{MnSymbolE}{}
\DeclareSymbolFont{MnLargeSymbols}{OMX}{MnSymbolE}{m}{n}
\SetSymbolFont{MnLargeSymbols}{bold}{OMX}{MnSymbolE}{b}{n}
\DeclareFontShape{OMX}{MnSymbolE}{m}{n}{
    <-6>  MnSymbolE5
   <6-7>  MnSymbolE6
   <7-8>  MnSymbolE7
   <8-9>  MnSymbolE8
   <9-10> MnSymbolE9
  <10-12> MnSymbolE10
  <12->   MnSymbolE12
}{}
\DeclareFontShape{OMX}{MnSymbolE}{b}{n}{
    <-6>  MnSymbolE-Bold5
   <6-7>  MnSymbolE-Bold6
   <7-8>  MnSymbolE-Bold7
   <8-9>  MnSymbolE-Bold8
   <9-10> MnSymbolE-Bold9
  <10-12> MnSymbolE-Bold10
  <12->   MnSymbolE-Bold12
}{}

\let\llangle\@undefined
\let\rrangle\@undefined
\DeclareMathDelimiter{\llangle}{\mathopen}%
                     {MnLargeSymbols}{'164}{MnLargeSymbols}{'164}
\DeclareMathDelimiter{\rrangle}{\mathclose}%
                     {MnLargeSymbols}{'171}{MnLargeSymbols}{'171}
\makeatother

 \newcommand{\beq}{\begin{equation}}
                \newcommand{\bea}{\begin{eqnarray}}
                \newcommand{\eea}{\end{eqnarray}}
                 \newcommand{\eeq}{\end{equation}}

\newcommand{\iM}{{\mathscr M}}

\newcommand{\mM}{{\mathfrak M}}

\newcommand {\BC}   {\mathbb C}

\newcommand {\BN}   {\mathbb N}

\newcommand {\BR}   {\mathbb R}
\newcommand {\BP}   {\mathbb P}

\newcommand {\bN}   {\mathbf{N}}
\newcommand {\bM}   {\mathbf{M}}
\newcommand {\bY}   {\mathbf{Y}}

\newcommand {\bQ}   {\mathbf{Q}}

\newcommand {\qe} {\mathfrak q}

\newcommand {\ib} {\mathbf{i}}

\newcommand {\jb} {\mathbf{j}}

\newcommand {\Hf} {\mathsf{H}}

\newcommand {\bT}   {\mathbf{T}}
\newcommand {\ba}  {\underline{\ac}}
\newcommand {\bb}  {\underline{\fb}}
\newcommand {\bqe} {\underline{\qe}}
\newcommand {\cb} {\mathbf{c}}

\newcommand {\bn}{\underline{\mathbf{n}}}

\newcommand {\bv} {\underline{\mathbf{v}}}

\newcommand {\bw} {\underline{\mathbf{w}}}

\newcommand {\bnu} {{\boldsymbol{\nu}}}
\newcommand {\bro} {{\boldsymbol{\rho}}}

\newcommand {\bla} {\underline{\boldsymbol{\lambda}}}

\newcommand {\bom} {{\boldsymbol{\varpi}}}

\newcommand {\BZ}   {\mathbb Z}
\newcommand {\ac} {\mathfrak{a}}
\newcommand {\fb} {\mathfrak{b}}

\newcommand {\CalB} {\mathcal B}

\newcommand {\CalE} {\mathcal E}

\newcommand {\CalI} {\mathcal I}

\newcommand {\CalK} {\mathcal K}
\newcommand {\CalL} {\mathcal L}

\newcommand {\CalN} {\mathcal N}

\newcommand {\CalP} {\mathcal P}

\newcommand {\CalR} {\mathcal R}
\newcommand {\CalS} {\mathcal S}
\newcommand {\CalT} {\mathcal T}

\newcommand {\CalV} {\mathcal V}
\newcommand {\CalX} {\mathcal X}

\newcommand {\CalW} {\mathcal W}
\newcommand {\CalZ} {\mathcal Z}
\newcommand{\ve}{\varepsilon}
\newcommand{\ep}{\epsilon}

\renewcommand{\hat}{\widehat}

\newcommand{\Mf}{\mathsf{M}}
\newcommand{\Gf}{\mathsf{G}}

\newcommand{\Gammadi}{\boldsymbol{\Gamma}}
\newcommand{\Gamab}{{\Gamma}_{\rm ab}}
\newcommand{\Gamav}{{\Gamma}^{\vee}_{\rm ab}}

\newcommand{\Vg}{\mathsf{Vert}_{\gamma}}

\newcommand{\Eg}{\mathsf{Edge}_{\gamma}}

\newcommand{\Tr}{\mathsf{Tr}\,}

\newcommand{\subsubsec}[1]{\subsubsection{\uwave{\sl #1}}}
\newcommand{\subsec}[1]{\subsection{\underline{\bf #1}}}
\newcommand{\secc}[1]{\section{$\mathbf{ #1}$}}

\begin{document}
\title[BPS/CFT, QQ-characters, Gauge origami]{BPS/CFT correspondence III:\\ \ \\  Gauge origami\\ partition function\\
and {\it qq}-characters}

\author{Nikita Nekrasov}

\address{Simons Center for Geometry and Physics\\
Stony Brook University, Stony Brook NY 11794-3636, USA
\\
Kharkevich Institute for Information Transmission Problems, Moscow 127051 Russia\\
E-mail: nikitastring@gmail.com}

\maketitle

\begin{abstract}
We study generalized gauge theories engineered by taking the low energy limit of the $Dp$ branes wrapping $X \times {\bT}^{p-3}$, with $X$ a possibly singular surface in a Calabi-Yau fourfold $Z$. For toric $Z$ and $X$ the partition function can be computed by localization, making it a statistical mechanical model, called the \emph{gauge origami}. The random variables are the ensembles of Young diagrams. The building block of the gauge origami is associated with a tetrahedron, whose edges are colored by vector spaces. We show the properly normalized partition function is an entire function of the Coulomb moduli, for generic values of the $\Omega$-background parameters. The orbifold version of the theory defines the $qq$-character operators, with and without the surface defects. 
The analytic properties are the consequence of a relative compactness of the moduli spaces ${\iM}({\vec n}, k)$
of crossed and spiked instantons, demonstrated in "BPS/CFT correspondence II: instantons at crossroads, moduli and compactness theorem". 

\end{abstract}

\setcounter{tocdepth}{2} 
\tableofcontents

\secc{ Introduction}\label{aba:sec1}

This paper is a continuation of the series \cite{Nekrasov:2015wsu, Nekrasov:2016qym}. 
There we proposed a set of observables in quiver ${\CalN}=2$ supersymmetric gauge theories. These observables
are useful in organizing the non-perturbative Dyson-Schwinger equations. The latter relate different instanton sectors contributions to the expectation values of gauge invariant chiral ring observables. We also introduced the geometric
setting to which these observables belong in a natural way. Namely, we defined the moduli spaces ${\mM}_{X, G}$ of what might be called supersymmetric gauge fields in the generalized gauge theories, whose space-time $X$ contains several, possibly intersecting, components:
\beq
X = \bigcup\limits_{A}\ X_{A}\ . 
\eeq
The gauge groups $G\vert_{X_{A}} = G_{A}$ on different components may be different. The intersections $X_{A} \cup X_{B}$ lead to the matter fields charged under the product group $G_{A} \times G_{B}$ (bi-fundamental multiplets).  In this paper we shall be studying the integrals over the moduli space ${\mM}_{X,G}$, which we shall compute using equivariant localization.

\subsec{Acknowledgments}

I am grateful to H.~Nakajima for patient explanations about the quiver varieties and to the anonymous referee for very useful comments on the manuscript. 
The results of this paper were reported at various seminars \cite{Nekrasov:IPPI},\cite{Nekrasov:SCGP}. 
The  research  was  carried  out  at  the  IITP RAS  with the support  of  the  Russian  Foundation  for  Sciences (project No. 14-50-00150).

\secc{Review\ of\ notations}\label{aba:sec2}

\subsec{Sets and partitions}

\subsubsection{Sequences}
For two sets $X$ and $S$ let $X^{S} = {\rm Maps}(S, X)$ denote the set of maps from $S$
to $X$. For a map $f: S \to X$ we sometimes use the notation
 \beq
 (x_{s})_{s\in S} \ ,
 \label{eq:mnot}
 \eeq with $x_{s} = f(s) \in X$. For example, a sequence $(a_{n})$, $n \in {\BN}$ would be denoted as $(a_{n})_{n \in {\BN}}$ or $(a_{n})_{n \geq 1}$, if the context is clear.

\subsubsection{Non-negative integers} 
are denoted by ${\BZ}_{\geq 0} = {\BN} \cup \{0 \}$. 

\subsubsection{Finite sets}
Let $[n]$ denote the set $\{ \, 1, 2, \ldots , n \, \}$ for $n \in {\BN}$.
For a finite set $X$ we denote by ${\#}X$ the number of its elements. Thus, for finite $X$ and $S$ 
\beq
{\#} X^{S} = \left( {\#}X \right)^{{\#}S}
\eeq

\subsubsection{Partitions}
There are lots of sums over partitions in this paper. 
Let $\Lambda$ denote the set of all partitions. 
An element ${\lambda} \in \Lambda$ is a non-increasing sequence ${\lambda} = ({\lambda}_{1} \geq {\lambda}_{2} \geq \ldots \geq {\lambda}_{{\ell}({\lambda})} > {\lambda}_{{\ell}({\lambda})+1} = {\lambda}_{{\ell}({\lambda})+2} = \ldots = 0  )$ of integers, with a finite number of positive terms, sometimes called {\it the parts of} ${\lambda}$. The number ${\ell}({\lambda})$ of positive terms is called the {\it length of the partition} $\lambda$, the sum
\beq
\sum_{i=1}^{{\ell}({\lambda})} {\lambda}_{i} = | {\lambda} |
\eeq
is called its {\it size}. We also identify the partitions $\lambda$ with the finite subsets of ${\BN}^{2} = {\BN} \times {\BN}$, as follows:
\beq
{\lambda} = \{ \ {\square} \ | \ {\square} = (i,j), \ i, j \geq 1, \ 1\leq j \leq {\lambda}_{i} \ \}
\label{eq:lamset}
\eeq  
The size $| {\lambda} |$ of the partition $\lambda$ is the number of elements ${\#}{\lambda}$ of the corresponding finite set. 
Not every finite subset of ${\BN}^{2}$ corresponds to a partition, only those, for which the complement ${\BN}^{2} \backslash {\lambda}$ is preserved by the action of the semi-group ${\BZ}_{\geq 0} \times {\BZ}_{\geq 0}$ on ${\BN}^{2}$ by translations. Equivalently, the partitions are in one-to-one correspondence with finite codimension monomial ideals in the ring of polynomials in two variables: ${\lambda} \leftrightarrow {\CalI}_{\lambda}$, ${\CalI}_{\lambda} \subset {\BC}[x,y]$, ${\CalI}_{\lambda} = \cup_{i=1}^{{\ell}_{\lambda}+1} \, {\BC}[x,y] x^{i-1}y^{{\lambda}_{i}}$.

We denote by ${\Lambda}[k]$ the set of partitions of $k$, i.e. the set of all ${\lambda} \in {\Lambda}$, such that $|{\lambda}| = k$. We have:
\beq
{\Lambda} = \bigsqcup\limits_{k \geq 0} \ {\Lambda}[k]
\label{eq:lambdec}
\eeq
The celebrated Euler formula:
\beq
\sum_{k=0}^{\infty} \ \#{\Lambda}[k] \, {\qe}^{k} = \prod_{n=1}^{\infty} \frac{1}{1-{\qe}^{n}}
\eeq
\subsec{Four and Six}

Let ${\4}$ denote the set $[4]$ and let
${\6}$ denote the set of $2$-element subsets of $\4$ (we write $ab$ instead of $\{ a, b \}$ to avoid the clutter):
\beq
{\4} =  \{ 1, 2, 3, 4 \} \, , \qquad {\6} = \left( \begin{matrix} {\4} \\ 2 \end{matrix} \right)  = \{ \, 12 ,\, 13 , \, 14 , \, 23 ,\,  24 ,\,  34 \, \} 
\label{eq:46sets} 
\eeq
In \eqref{eq:46sets} we exhibit the lexicographic order of the sets $\4$ and $\6$ which is used below in some formulas. For example, $12 < 14 < 23 < 34$. For $A \in {\6}$ we denote by ${\bar A} = {\4} \backslash A$ its complement. Let $\3$ denote the quotient ${\6}/{\sim}$ where $A \sim {\bar A}$. Identify ${\3} = \{ 1, 2, 3 \} \subset {\4}$ by choosing a representative $A = a4$, $a \in [3]$. Define the map: ${\varphi}: {\6} \to {\4}$ by
 \beq
{\varphi}(A) = {\rm inf} \, {\bar A}  \ \in {\4}, \qquad
\label{eq:gamap}
\eeq
so that 
\beq
 {\varphi}(12) = 3,\ {\varphi}(13) = {\varphi}(14) = 2, \   {\varphi}(24) = {\varphi}(23) = {\varphi}(34) = 1\ . 
\eeq
We also define the following map ${\ve}: {\6} \to {\BZ}_{2}$:
write $A = \{a , b\}$, $a < b \in {\4}$, write ${\bar A} = \{c , d\}$, $c < d \in {\4}$, then ${\ve}(A) = {\ve}_{abcd}$. Thus, 
\beq
 {\ve}(12) = {\ve} (34 ) = {\ve} ( 14 ) = {\ve} ( 23 ) = +1 , \ \
 {\ve} ( 13 ) = {\ve} ( 24 ) = - 1 \ . \
\eeq
It may seem surprising that ${\ve}$ takes values $+1$ four times and $-1$ only two times, but in fact it is natural, since ${\ve}(A) = {\ve}({\bar A})$, therefore $\ve$ is defined on $\3$. Since a two-valued function on a set of odd cardinality cannot split it equally, more classes
are bound to be good rather then bad (assuming the values $+1$ and $-1$ are identified with ``good'' and ``bad''). 

\medskip
\centerline{\picti{3}{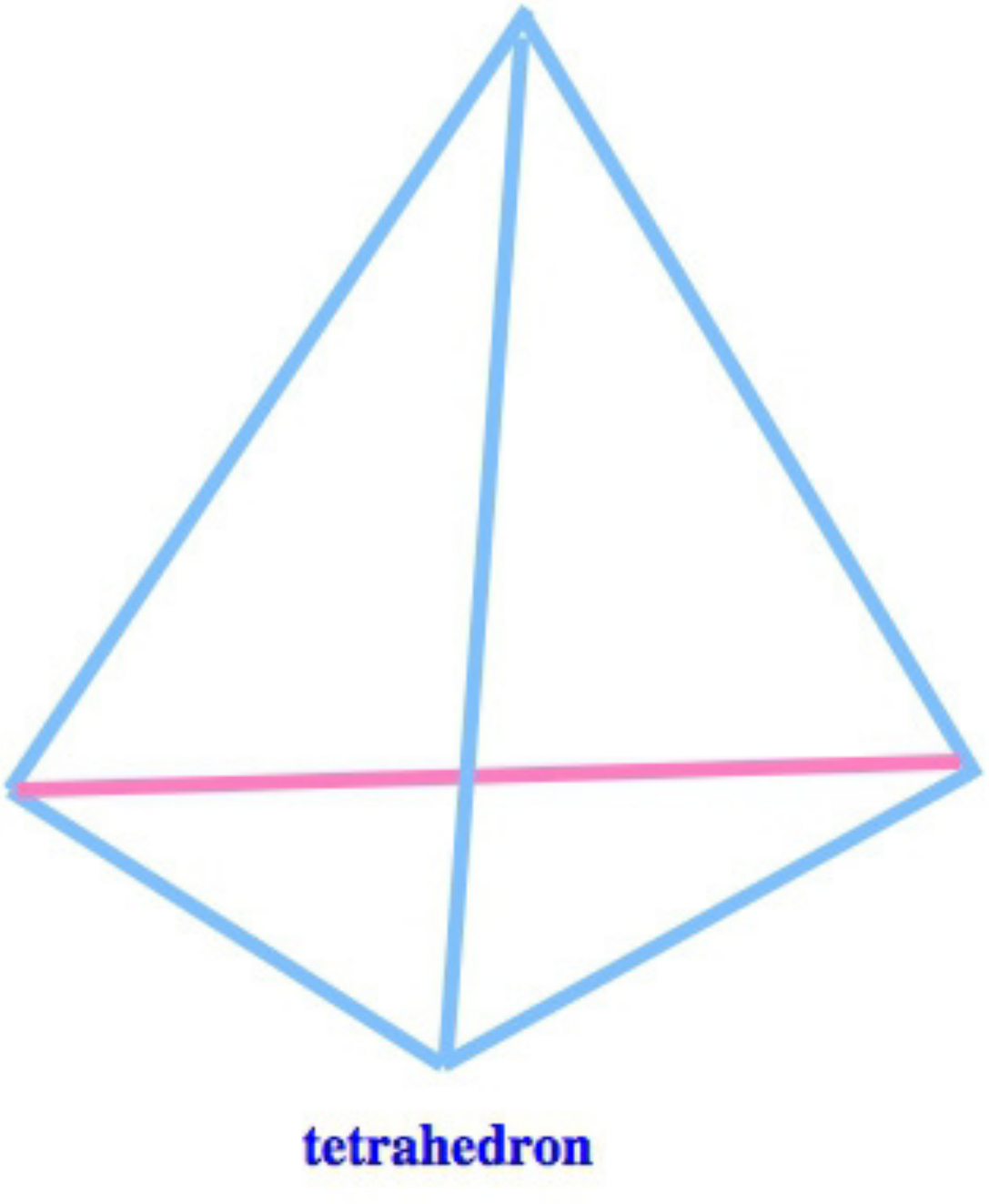}}
\medskip
\noindent 
It is useful to view $\4$ as the set of faces (or vertices) of the tetrahedron, while $\6$ is the set of edges. The edge $ab$ connects the vertices $a$ and $b$. Alternatively the edge $ab$ is the common boundary of the faces $a$ and $b$. 

\subsec{Finite groups and quiver varieties}

\subsubsec{Abelian groups}
We denote by $\Gamab$ a finite abelian group. It is well-known that any such $\Gamab$ is a product of cyclic groups whose orders are powers of primes:
\beq
{\Gamab} = \varprod_{\kappa = 1}^{d} \left( {\BZ}/p_{\kappa}^{l_{\kappa}} {\BZ} \right)  \, , \qquad l_{\kappa} \in {\BN}, \quad p_{\kappa} \, - \, {\rm primes}
\label{eq:finab}
\eeq
An element of $\Gamab$ is a string ${\bf t} = \left( t_{1}, \ldots , t_{d} \right)$ of integers
defined modulo lattice $t_{\kappa} \sim t_{\kappa} + p_{\kappa}^{l_{\kappa}} {\BZ}$. 
All irreducible representations ${\CalL}_{\varpi}$ of $\Gamab$ are complex one-dimensional,  labeled by a string of integers
\beq 
{\bnu} = \left( n^{1}, \ldots , n^{d} \right) \in {\Gamav}\, , \qquad n^{\kappa} \in {\BZ}
\eeq 
also 
defined modulo lattice $n^{\kappa} \sim n^{\kappa}+ p_{\kappa}^{l_{\kappa}} {\BZ}$:
\beq
T_{{\CalL}_{\bnu}} ( {\bf t} ) = {\exp} \ 2\pi\sqrt{-1} \,  \sum_{\kappa} \, \frac{t_{\kappa}n^{\kappa}}{p_{\kappa}^{l_{\kappa}}}
\label{eq:1char}
\eeq
We set ${\bnu} = \boldsymbol{0}$ to label the trivial representation with all $n^{\kappa} = 0$, 
\beq
T_{{\CalL}_{\boldsymbol{0}}}({\bf t}) \equiv 1
\eeq 
The set $\Gamav$ is also an abelian group, isomorphic to $\Gamab$, with multiplication given by the tensor product of irreducible representations.  We shall be using the addition symbol for the group law on ${\Gamav}$:
\beq
{\CalL}_{{\bnu}_{1}+{\bnu}_{2}} =  {\CalL}_{{\bnu}_{1}}\otimes {\CalL}_{{\bnu}_{2}}\, , \qquad {\CalL}_{\bnu}^{*} = {\CalL}_{-{\bnu}}
\eeq
Let \beq
{\delta}_{\Gamav} : {\Gamav} \to \{ 0 , 1 \}
\eeq
be the indicator function of the trivial representation: 
\beq
{\delta}_{\Gamav} ({\boldsymbol{0}}) = 1, \qquad {\delta}_{\Gamav} ({\bnu}) = 0, \qquad {\bnu} \neq {\boldsymbol{0}}
\eeq

\subsubsec{Nonabelian subgroups of $SU(2)$}

Let ${\gamma}$ denote the affine Dynkin diagram of type $D$, or $E$, respectively (see the Fig. $1$):

\centerline{\picti{5}{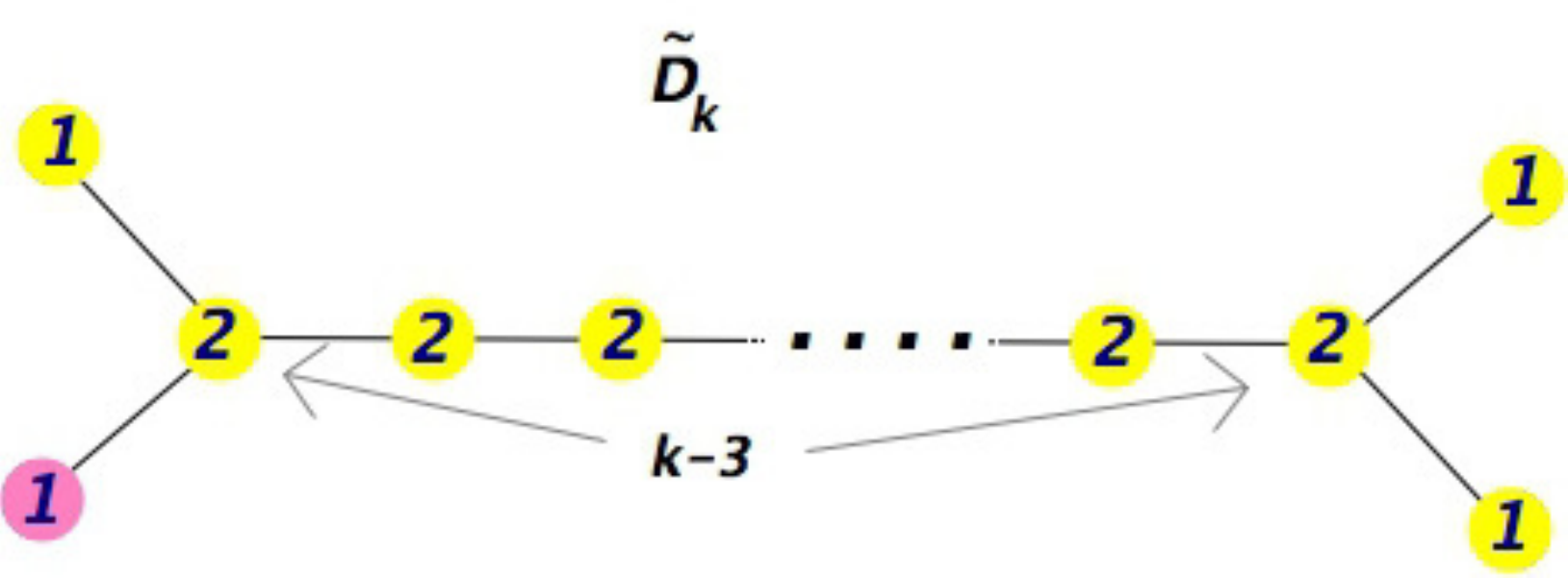}\qquad ' \picti{8}{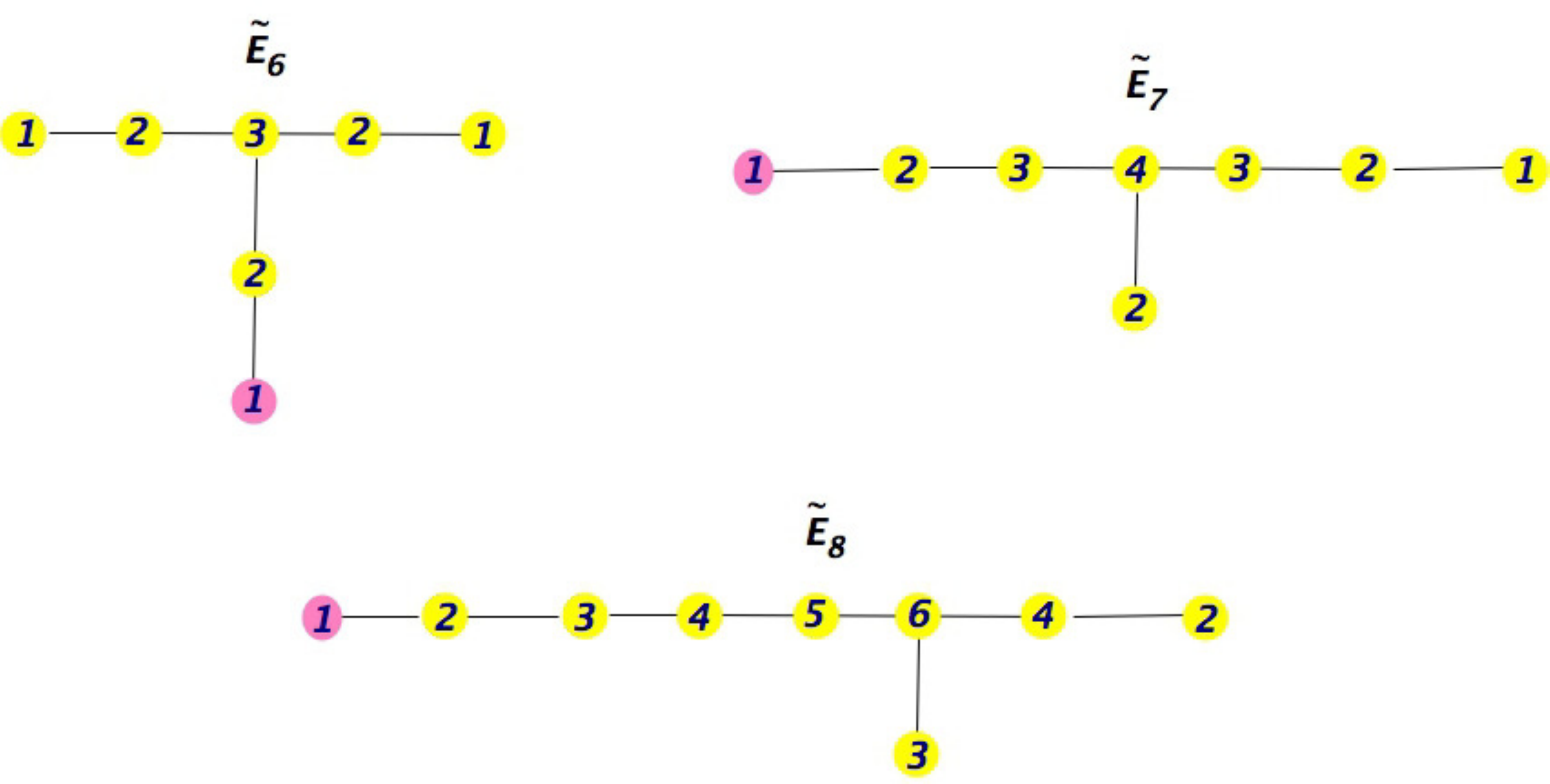}}
\centerline{{\bf Fig.}1}
\bigskip

Let ${\Vg}$ be the set of vertices of $\gamma$, 
${\Eg}$ be the set of oriented edges (we pick any orientation). For
the edge $e \in {\Eg}$ let $s(e), t(e) \in {\Vg}$ denote
its source and target, respectively. 

Let ${\Gamma}_{\gamma} \subset SU(2)$ denote the corresponding
non-abelian finite subgroup. For ${\gamma} = {\tilde E}_{6, 7, 8}$ the group $\Gamma_{\gamma}$ is the binary tetrahedral, octahedral, icosahedral group, respectively. 
 
In this correspondence ${\ib} \in {\Vg}$ labels the 
irreducible representations $R_{\ib} \in {\Gamma}^{\vee}_{\gamma}$ of ${\Gamma}_{\gamma}$. The edges $\Eg$ show up in the tensor products: let ${\bf 2}$ denote the defining two-dimensional representation of $SU(2)$. Then:
\beq
{\bf 2} \otimes R_{\ib} = \bigoplus_{e \in s^{-1}({\ib})} \, R_{t(e)} \ \oplus \  \bigoplus_{e \in t^{-1}({\ib})} \, R_{s(e)}
\label{eq:mckay}
\eeq
where ${\bf 2}$ is viewed as the representation of ${\Gamma}_{\gamma} \subset SU(2)$. 
The dimensions ${\rm dim}R_{\ib}$ are indicated on the corresponding nodes in the picture, the vector of dimensions is annihilated by the affine Cartan matrix $ = 2 - $incidence matrix of $\gamma$, cf. \eqref{eq:mckay}:
\beq
2 {\rm dim} R_{\ib} = \sum_{e \in s^{-1}({\ib})} \, {\rm dim}R_{t(e)} \ + \  \sum_{e \in t^{-1}({\ib})} \, {\rm dim}R_{s(e)}
\label{eq:mckay2}
\eeq  The trivial representation is colored pink on Fig. 1. 

\subsubsec{Walks on quivers}

Let ${\ib}_{s}, {\ib}_{t} \in {\Vg}$. 
A path $p$ connecting ${\ib}_{s}$ (the source of $p$) to
${\ib}_{t}$ (the target of $p$) of length ${\ell}_{p}$
on the quiver 
$\gamma$ is the ordered sequence of pairs
$p_{i} = (e_{i}, {\sigma}_{i})$, $i = 1, \ldots , {\ell}_{p}$   
where $e_{i} \in {\Eg}$, ${\sigma}_{i} = \pm 1$, and
\begin{enumerate}
\item the source of $p$: if ${\sigma}_{1} = 1$, then $s(e_{1}) = {\ib}_{s}$, otherwise $t(e_{1}) = {\ib}_{s}$
\item the end-point of $p$: if ${\sigma}_{{\ell}_{p}} = 1$, then 
$t(e_{{\ell}_{p}}) = {\ib}_{t}$, otherwise $s(e_{{\ell}_{p}}) = {\ib}_{t}$
\item concatenation: if ${\sigma}_{i} = 1$, ${\sigma}_{i+1} = 1$, then 
$t(e_{i}) = s(e_{i+1})$, if ${\sigma}_{i} = 1$, ${\sigma}_{i+1} = -1$, then 
$t(e_{i}) = t(e_{i+1})$, if ${\sigma}_{i} = -1$, ${\sigma}_{i+1} = 1$, then 
$s(e_{i}) = s(e_{i+1})$, if ${\sigma}_{i} = -1$, ${\sigma}_{i+1} = -1$, then 
$s(e_{i}) = t(e_{i+1})$
\end{enumerate}
Let us denote the set of all paths on $\gamma$ connecting ${\ib}_{s}$ to ${\ib}_{t}$
by ${\CalP}_{{\ib}_{s}}^{{\ib}_{t}}[{\gamma}]$. There is an obvious associative concatenation
map:
\begin{multline}
\star \, : \, {\CalP}_{{\ib}_{1}}^{{\ib}_{2}}[{\gamma}] \times 
{\CalP}_{{\ib}_{2}}^{{\ib}_{3}}[{\gamma}] \, \longrightarrow \, {\CalP}_{{\ib}_{1}}^{{\ib}_{3}}[{\gamma}] \ , \\
p \times {\tilde p} \mapsto {\tilde p} \star p\, , \qquad
\left( {\tilde p} \star p  \right)_{i} = \Biggl\{ \begin{matrix} & \ p_{i} \ , & \ 1 \leq i \leq {\ell}_{p'}  \\
& \ {\tilde p}_{i-{\ell}_{p}} \  , & \ {\ell}_{p} < i \leq {\ell}_{p} + {\ell}_{\tilde p} = {\ell}_{{\tilde p} \star p} \end{matrix}
\end{multline}
and the inversion map 
\begin{multline}
- \, : \, {\CalP}_{{\ib}_{s}}^{{\ib}_{t}}[{\gamma}] \, \longrightarrow \, {\CalP}_{{\ib}_{t}}^{{\ib}_{s}}[{\gamma}] \ , \\
p \mapsto {\bar p}  , \qquad
{\bar p}_{i} = \left( e_{{\ell}_{p}+1-i} , - {\sigma}_{{\ell}_{p}+1-i} \right)  \ ,  \ 1 \leq i \leq {\ell}_{p}
\end{multline}

\subsubsec{Nakajima varieties}

Define the Nakajima varieties ${\mM}_{\gamma}({\bv}, {\bw})$   associated with a quiver $\gamma$ and two dimension vectors ${\bv}, {\bw} \in {\BZ}^{\Vg}_{\geq 0}$ \cite{Nakajima:1994, Nakajima:1994r, Nakajima:1998}. 

Let $\gamma$ be as before. To each vertex $\ib \in \Vg$ we associate two Hermitian vector spaces $W_{\ib}$, $V_{\ib}$ of dimensions $w_{\ib}, v_{\ib}$, respectively. Let 
\beq
U_{\gamma} ({\bv}) = \varprod_{\ib \in {\rm Vert}_{\gamma}}\ U(v_{\ib})
\label{eq:ubv}
\eeq
be the group of unitary transformations of ${\bf V} = ( V_{\ib} )_{{\ib}\in {\Vg}}$. 
First, form the Hermitian vector space:
\begin{multline}
H_{\gamma} ({\bv}, {\bw}) = T^{*} \left( \, \bigoplus_{e \in {\rm Edge}_{\gamma}} {\rm Hom} (V_{s(e)}, V_{t(e)})  \ \bigoplus \ \bigoplus_{{\ib} \in {\Vg}} \, {\rm Hom} (W_{\ib}, V_{\ib}) \, \right) = \\
\Biggl\{ \, \left( B_{e}, {\tilde B}_{e} \right)_{e \in {\rm Edge}_{\gamma}}, \ \left( I_{\ib}, J_{\ib} \right)_{\ib \in \Vg} \, \Biggr\vert \, I_{\ib}: W_{\ib} \to V_{\ib}\, , \ J_{\ib}: V_{\ib} \to W_{\ib} \, , \, B_{e} : V_{s(e)} \to V_{t(e)} \, , {\tilde B}_{e} : V_{t(e)} \to V_{s(e)} \, \Biggr\} \label{eq:adhmnak}
\end{multline}
which is acted upon by $U_{\gamma} ({\bv})$ via:
\beq
\left( u_{\omega} \right)_{\omega \in {\rm Vert}_{\gamma}} \cdot \left( \left( B_{e}, {\tilde B}_{e} \right)_{e \in {\rm Edge}_{\gamma}}, \ \left( I_{\ib}, J_{\ib} \right)_{\ib \in \Vg} \, \right) = \left( \left( u_{t(e)} B_{e} u_{s(e)}^{-1}, u_{s(e)} {\tilde B}_{e} u_{t(e)}^{-1} \right)_{e \in {\rm Edge}_{\gamma}} \, , \ \left( u_{\ib} I_{\ib}, J_{\ib} u_{\ib}^{-1} \right)_{\ib \in \Vg}  \, \right)
\label{eq:gaga}
\eeq
For a path $p \in {\CalP}_{{\ib}_{t}}^{{\ib}_{s}}[{\gamma}]$ define its holonomy ${\CalB}_{p}: V_{{\ib}_{s}} \to V_{{\ib}_{t}}$ in the obvious way: 
\beq
{\CalB}_{p} = \prod_{i=1}^{\atop{\leftarrow}{{\ell}_{p}}}
\, \Biggl\{  \begin{matrix}
 B_{e_{i}}\ , \quad {\sigma}_{i} = +1 \\
 {\tilde B}_{e_{i}}\ , \quad {\sigma}_{i} = - 1\end{matrix}
\label{eq:bmon}
\eeq
This definition is compatible with the path multiplication:
\beq
{\CalB}_{p_{2}} {\CalB}_{p_{1}} = {\CalB}_{p_{2}\star p_{1}}
\eeq
The action \eqref{eq:gaga} preserves the hyper-K\"ahler structure of $H_{\gamma} ({\bv}, {\bw})$, with
the three symplectic forms ${\varpi}_{I, J, K}$ given by:
\begin{multline}
{\varpi}_{I} = \sum_{e \in {\Eg}} {\Tr}_{V_{t(e)}} \left(  dB_{e} \wedge dB_{e}^{\dagger} - d{\tilde B}_{e}^{\dagger} \wedge d{\tilde B}_{e} \right) +  \sum_{{\ib} \in {\Vg}} {\Tr}_{W_{\ib}}\, \left( dJ_{\ib} \wedge dJ^{\dagger}_{\ib} - dI^{\dagger}_{\ib} \wedge dI_{\ib} \right) \, , \\
{\varpi}_{J} + \sqrt{-1} {\varpi}_{K} = \sum_{{\ib} \in {\Vg}} {\Tr}_{W_{\ib}}\, \left( dJ_{\ib} \wedge dI_{\ib} \right) + \sum_{e \in {\Eg}} {\Tr}_{V_{t(e)}} \left(  dB_{e} \wedge  d{\tilde B}_{e} \right) 
\end{multline}
Then perform the hyper-K\"ahler reduction with respect to the action \eqref{eq:gaga}:
\beq
{\mM}_{\gamma} ({\bv} ,{\bw}) = {\vec\mu}^{-1} ({\vec\zeta}) /U_{\gamma}({\bv})
\label{eq:hkq}
\eeq
where ${\vec\mu} = ({\mu}_{I, \ib}, {\mu}_{J, \ib}, {\mu}_{K, \ib})_{\ib \in \Vg}$, 
\begin{multline}
{\mu}_{I, \ib} \, = \, I_{\ib} I_{\ib}^{\dagger} - J_{\ib}^{\dagger} J_{\ib}  \, + \ \sum_{e \in t^{-1}({\ib})} \, \left( B_{e} B_{e}^{\dagger} - {\tilde B}_{e}^{\dagger} {\tilde B}_{e} \right) \ + \ \sum_{e \in s^{-1}({\ib})}\, \left(  {\tilde B}_{e} {\tilde B}_{e}^{\dagger} - B_{e}^{\dagger} B_{e} 
\right)  \ , \\
{\mu}_{J, \ib} + \sqrt{-1} {\mu}_{K, \ib} \, = \, I_{\ib} J_{\ib} + \sum_{e \in t^{-1}({\ib})} B_{e} {\tilde B}_{e} - \sum_{e \in s^{-1}({\ib})} {\tilde B}_{e}  B_{e} 
\ ,
\end{multline}
and we take (this is not the most general definition)
\beq
{\vec\zeta} = ({\zeta}_{\ib} {\bf 1}_{V_{\ib}}, 0 , 0)_{\ib \in \Vg}
\label{eq:lev}
\eeq 
with all 
$\zeta_{\ib} > 0$. 

$\underline{Stability}$. 
Instead of solving three equations ${\vec\mu} = {\vec\zeta}$ one can actually solve only ${\mu}_{\BC} \equiv {\mu}_{J} + \sqrt{-1} {\mu}_{K} = 0$, and then take a quotient of the set of stable points in ${\mu}_{\BC}^{-1}(0)$ by the action of 
\beq
G_{\gamma} ({\bv}) = \varprod_{\ib \in {\Vg}}\ GL(v_{\ib}; {\BC})
\label{eq:gbv}
\eeq
so that
\beq
{\mM}_{\gamma} ({\bv}, {\bw}) = {\mu}_{\BC}^{-1}(0)^{\rm stable}/ G_{\gamma}({\bv})
\label{eq:stb}
\eeq
The stable points are the $G_{\gamma}({\bv})$-orbits of $(B_{e}, {\tilde B}_{e}, I, J)$ s.t.
the path algebra of $\gamma$ represented by the products of $B_{e}$ and ${\tilde B}_{e}$ 
acting on the image $\bigoplus_{\ib \in \Vg} I_{\ib}W_{\ib}$ generates all of $\bigoplus_{\ib\in \Vg} V_{\ib}$:
\beq
V_{\ib} = \sum_{{\ib}' \in \Vg} \sum_{p \in {\CalP}_{{\ib}'}^{\ib}[{\gamma}]} \ {\CalB}_{p}  \, I_{\ib'} W_{\ib'}\ .
\label{eq:vib}
\eeq
In other words: any collection ${\bf V}' = (V_{\ib}')_{{\ib} \in {\Vg}} \subset {\bf V}$ of vector subspaces $V_{\ib}' \subset V_{\ib}$, obeying: 
\beq
\begin{aligned}
& {\rm S1)} \qquad I_{\ib}W_{\ib} \subset V_{{\ib}}', \ {\rm for\ all}\ \ib \in \Vg \, , \\
&  {\rm S2)}\qquad  B_{e}(V_{s(e)}') \subset V_{t(e)}', \, {\tilde B}_{e}(V_{t(e)}') \subset V_{s(e)}' \,  , {\rm  for\ all}\ e \in \Eg \, \\
\end{aligned}
\label{eq:vv'}
\eeq must coincide with ${\bf V}$:
$V_{\ib}' = V_{\ib}$ for all ${\ib} \in {\Vg}$. 

A simple proof of the equivalence of \eqref{eq:hkq} and \eqref{eq:stb} can be found along the lines of the arguments of the section 3.4 and \cite{Nekrasov:2016qym}: in one direction, 
any solution to ${\mu}_{I, \ib} = {\zeta}_{\ib} \cdot {\bf 1}_{V_{\ib}}$ is stable. Indeed, ${\bf V}' \subset {\bf V}$ as in \eqref{eq:vv'}, and let $P_{\ib}$ denote the orthogonal projection $V_{\ib} \to V_{\ib}^{'\perp}$. By \eqref{eq:vv'} we have:
\beq
P_{\ib} I_{\ib} = 0 \, , \qquad P_{t(e)} B_{e} (1-P_{s(e)}) = 0 \, ,  \qquad P_{s(e)} {\tilde B}_{e} (1-P_{t(e)}) = 0 \eeq
Define $b_{e} = P_{t(e)}B_{e}P_{s(e)}$, ${\tilde b}_{e} = P_{s(e)}{\tilde B}_{e} P_{t(e)}$, 
$b_{e}'= (1-P_{t(e)})B_{e}P_{s(e)}$, ${\tilde b}_{e}'= (1-P_{s(e)}){\tilde B}_{e}P_{t(e)}$. 
Then 
\begin{multline}
{\zeta}_{\ib} {\rm dim}\left( V_{\ib}/V_{\ib}' \right) = {\Tr}_{V_{\ib}} \left( P_{\ib} {\mu}_{\ib} P_{\ib} \right) \, = \\
{\Tr}_{(V_{\ib}')^{\perp}} \left( - j_{\ib}^{\dagger} j_{\ib}  \, + \ \sum_{e \in t^{-1}({\ib})} \, \left( b_{e} b_{e}^{\dagger}  - {\tilde b}_{e}^{\dagger} {\tilde b}_{e} - {\tilde b}_{e}^{'\dagger} {\tilde b}'_{e}\right) \ + \ \sum_{e \in s^{-1}({\ib})}\, \left(  {\tilde b}_{e} {\tilde b}_{e}^{\dagger} - b_{e}^{\dagger} b_{e}  - b_{e}^{'\dagger}b'_{e}
\right) \right) \, ,
\end{multline}
hence, after obvious cancellations,
\beq
0 \leq \sum_{{\ib} \in \Vg} {\zeta}_{\ib} {\rm dim}\left( V_{\ib}/V_{\ib}' \right) = - \left( \sum_{\ib \in \Vg} {\Tr}_{V_{\ib}^{'\perp}}  j_{\ib}^{\dagger} j_{\ib}  \, + \ \sum_{e \in \Eg} \, {\Tr}_{V_{t(e)}^{'\perp}}  {\tilde b}_{e}^{'\dagger} {\tilde b}'_{e} \ + \ {\Tr}_{V_{s(e)}^{'\perp}}\,  b_{e}^{'\dagger}b'_{e}
 \right) \leq 0
\eeq 
which implies $V_{\ib}' = V_{\ib}$ for all $\ib \in \Vg$. 
Conversely, given a stable solution $(B_{e}, {\tilde B}_{e}, I_{\ib}, J_{\ib})$ to ${\mu}_{\BC} = 0$ equations, run the gradient flow of the function:
\beq
f = \frac 12 \sum_{\ib \in \Vg} {\Tr}_{V_{\ib}} \left( {\mu}_{I, \ib} - {\zeta}_{\ib} {\bf 1}_{V_{\ib}} \right)^{2}
\eeq
which goes along the $\times_{\ib} GL(V_{\ib})$ orbits. The end-point of the flow is either at $f = 0$ which would establish the rest of the equations in \eqref{eq:hkq}, or at the higher critical point. There, the $End(V_{\ib})$-matrices $h_{\ib} = {\mu}_{I,{\ib}} - {\zeta}_{\ib} {\bf 1}_{V_{\ib}}$ solve:
\beq
h_{t(e)}B_{e} = B_{e} h_{s(e)}\, , \qquad h_{s(e)}{\tilde B}_{e} = {\tilde B}_{e} h_{t(e)}\, , \qquad
 h_{\ib} I_{\ib} = 0 \, , \ J_{\ib} h_{\ib} = 0 
 \label{eq:hbij}
 \eeq
Therefore $V_{\ib}' = {\rm ker}h_{\ib}$ obeys both S1) and S2) conditions of \eqref{eq:vv'}, therefore $h_{\ib} =0$ for all $\ib \in \Vg$. 

\subsubsec{Framing\ symmetries\ of\ Nakajima\ varieties} The Nakajima variety ${\mM}_{\gamma} ({\bv}, {\bw})$  has a symmetry group 
\beq
U_{\gamma}({\bw}) = \varprod\limits_{\ib \in {\Vg}} U(w_{\ib}) \ 
\eeq acting in an obvious way on the operators $(I_{\ib}, J_{\ib})$.  The maximal torus $T_{\gamma}({\bw}) \subset U_{\gamma}({\bw})$ fixed point locus 
is the union
\beq
{\mM}_{\gamma} ({\bv}, {\bw})^{T_{\gamma}({\bw})} = \bigsqcup_{{\bv} \, = \, \sum\limits_{{\ib} \in {\Vg}, \, {\alpha} \in [w_{\ib}]} \ {\bv}^{\ib, \alpha}}\qquad  \varprod\limits_{\ib \in\Vg} \varprod\limits_{{\alpha} \in [w_{\ib}]} {\mM}_{\gamma} ({\bv}^{\ib, \alpha}, {\underline{\delta}}_{\ib})
\label{eq:fundec}
\eeq
where ${\bv}^{\ib, \alpha} \in {\BZ}^{\Vg}_{\geq 0}$ for each $({\ib}, {\alpha})$, i.e.
\beq
{\bv}^{\ib, \alpha} = \left( v^{\ib, \alpha}_{\tilde\ib} \right)_{{\tilde\ib} \in {\Vg}} \, , 
\label{eq:locdimvec}
\eeq
and 
\beq
{\underline{\delta}}_{\ib} = ( {\delta}_{\ib, \jb} )_{\jb \in \Vg}
\eeq
We define the {\emph{fundamental}} Nakajima variety 
\beq
{\bM}^{\ib}_{\gamma}({\bv}) = {\mM}_{\gamma} ({\bv}, {\underline{\delta}}_{\ib})
\label{eq:fnv}
\eeq
The Eq. \eqref{eq:fundec} explains the importance of the fundamental Nakajima varieties. 

\subsubsec{Nakajima-Young varieties}

The Nakajima varieties ${\mM}_{\gamma}({\bv}, {\bw})$ with the choice
\eqref{eq:lev} have a holomorphic ${\BC}^{\times}$-symmetry (its compact subgroup $U(1)$ acts by an isometry): $u \in {\BC}^{\times}$ acts via
\beq
u  \cdot \left( B_{e}, \, {\tilde B}_{e} \, , \, I_{\ib}, \, J_{\ib} \, \right)  \ = \ \left( u\, B_{e}, \ u\, {\tilde B}_{e} \, , \ u\, I_{\ib}, \, u\, J_{\ib} \, \right)
\label{eq:sac}
\eeq
Define Nakajima-Young variety
${\bY}_{\gamma}^{\ib} ({\mu})$ to be the connected component of the
fixed point set:
\beq
{\bM}_{\gamma}^{\ib} ({\bv})^{{\BC}^{\times}} = \bigsqcup_{{\mu} \in {\Lambda}_{\gamma}^{\ib} [{\bv}]} \, {\bY}_{\gamma}^{\ib} ( {\mu} )
\eeq
with 
\beq
{\Lambda}_{\gamma}^{\ib} [{\bv}] = {\pi}_{0}\left( {\bM}_{\gamma}^{\ib} ({\bv})^{{\BC}^{\times}}  \right)
\eeq denoting the set of connected components. 
We define the sets ${\Lambda}^{\ib}_{\gamma}$ for ${\ib} \in \Vg$:
\beq
{\Lambda}^{\ib}_{\gamma}  \ = \ \bigsqcup\limits_{{\bv} \in {\BZ}_{\geq 0}^{\Vg}} \ 
{\Lambda}^{\ib}_{\gamma} \left[ {\bv} \right]
\label{eq:nycc}
\eeq
For $\mu \in {\Lambda}^{\ib}_{\gamma} \left[ {\bv} \right]$ we define:
\beq
| {\mu} | = {\bv} \in {\BZ}_{\geq 0}^{\Vg}
\eeq
Each Nakajima-Young variety ${\bY}_{\gamma}^{\ib} ({\mu})$ carries a set of vector bundles:
\beq
V_{{\jb}, n}^{\ib} ({\mu}) \longrightarrow {\bY}_{\gamma}^{\ib} ({\mu})
\eeq
where ${\jb} \in \Vg$, $n \geq 0$, and
\beq
V_{{\jb}, n}^{\ib} ({\mu})
 \ = \sum_{p \in {\CalP}_{{\ib}}^{\jb}[{\gamma}], \ {\ell}_{p} = n}\ {\CalB}_{p} \, I ({\BC}) \ .
 \label{eq:gradvii}
 \eeq 
 The stability condition \eqref{eq:vib} implies, for any $\jb \in \Vg$:
 \beq
 V_{\jb} = \bigoplus_{n=0}^{\infty} \ V_{\jb, n}^{\ib}
 \label{eq:gradv}
 \eeq
 It is easy to show that $J \equiv 0$ on all $ {\bY}_{\gamma}^{\ib} ({\bv}; {\mu})$, and $V_{\jb,0}^{\ib} = {\BC} {\delta}_{\ib , \jb}$. Let us clarify the origin of the direct sum decomposition 
 \eqref{eq:gradv}. The ${\BC}^{\times}$-invariance of the $G_{\gamma}({\bv})$-orbit of
 $(B_{e}, {\tilde B}_{e}, I, J)$ means that the transformation \eqref{eq:sac} can be compensated by an element $(g_{\jb}(u))_{\jb \in \Vg}$:
 \beq
 g_{t(e)} (u) B_{e} g_{s(e)}(u)^{-1} = u\, B_{e}\, , \quad g_{s(e)} (s) {\tilde B}_{e} g_{t(e)}^{-1}(u) = u\, {\tilde B}_{e}\, , \qquad g_{\ib}(u) I = u\, I\, , \qquad J g_{\ib}(u)^{-1}  = u\, J \label{eq:bijs}
 \eeq
Then 
\beq
V_{\jb, n}^{\ib} = {\rm Ker}\, \left( g_{\jb}(u) - u^{n+1} \right)  \subset V_{\jb}
\eeq 
are obviously mutually orthogonal for different $n$'s. 
The ranks ${\nu}_{\jb, n}^{\ib} ({\mu}) = 
{\rm rk}V_{{\jb}, n}^{\ib} ({\mu})$
are important local invariants of ${\bY}_{\gamma}^{\ib} ({\mu})$. By definition:
\beq
\sum_{n=0}^{\infty} {\nu}_{\jb, n}^{\ib} ({\mu})  = v_{\jb}
\eeq
The $K$-theory class of the tangent bundle ${\CalT}_{{\bY}_{\gamma}^{\ib} ({\mu})}$ to ${\bY}_{\gamma}^{\ib} ({\mu})$ can be expressed in terms of those of $V_{\jb, n}^{\ib}$:
\beq
\begin{aligned}
& \left[ {\CalT}_{{\bY}_{\gamma}^{\ib} ({\mu})} \right] \ = \ \left[  V_{\ib , 0}^{\ib} \right]  \\
& \qquad\qquad\qquad\ +\ \sum\limits_{n \geq 0, \, e \in \Eg} \  \left[  {\rm Hom}\left( V_{t(e),n}^{\ib} , V_{s(e),n+1}^{\ib} \right) \oplus {\rm Hom}\left( V_{s(e),n}^{\ib} , V_{t(e),n+1}^{\ib} \right) \right]  \\
& \qquad\qquad\qquad\qquad\ -\  \sum\limits_{n \geq 0, \, {\jb} \in \Vg}  \left[  {\rm Hom}\left( V_{{\jb},n}^{\ib} , V_{{\jb},n}^{\ib} \right) \oplus {\rm Hom}\left( V_{{\jb},n}^{\ib} , V_{{\jb},n+2}^{\ib} \right)\right] \  .\\
\end{aligned}
\label{eq:tny}
\eeq

{}$\underline{Remark}.$ In the case of $\gamma = {\hat A}_{0}$, where $\Eg = \{ e \}$, $\Vg = \{ 0 \}$, $s(e) = t(e) = 0$, the fundamental Nakajima variety is the Hilbert scheme of $v$ points on ${\BC}^{2}$, a.k.a. the moduli space of noncommutative $U(1)$ instantons on ${\BR}^{4}$, while the Nakajima-Young varieties are the connected components of the so-called graded Hilbert scheme of $v$ points.  

\subsec{The local model data}

To specify the basic local model data we fix:
\begin{enumerate}
\item{}
The string 
\beq
{\bar\ve}  = ( {\ve}_{a} )_{a \in {\4}}
\label{eq:vep}
\eeq
 of $4$ complex numbers which sum to zero:
\beq
{\ve}_{1} + {\ve}_{2} + {\ve}_{3} + {\ve}_{4} = 0 
\label{eq:epar}
\eeq

\item{}
The string $\bar n$ of 
$6$ non-negative integers $n_{A} \geq 0$, $A \in {\6}$. Let
\beq
{\bN} = \bigsqcup_{A \in {\6}} [n_{A}] \approx \{ (A, {\alpha} ) \, | \, A \in {\6}, {\alpha} \in [n_{A}] \}\ . 
\label{eq:nlab}
\eeq

\item{} 
The string ${\bar\ac} \in {\BC}^{\bN}$ of 
$$
\sum_{A \in {\6}} n_{A}
$$ complex numbers ${\ac}_{A, {\alpha}} \in {\BC}$, ${\alpha} = 1, \ldots , n_{A}$, also denoted as 
\beq
{\bf\ac}_{A} = \left( {\ac}_{A, {\alpha}} \right)_{{\alpha} \in [n_{A}]} \equiv \left( {\ac}_{A, 1} , \ldots , {\ac}_{A, n_{A}} \right) \in {\BC}^{n_{A}}\ .
\label{eq:avec}
\eeq 

\item{}
The fugacity 
\beq
{\qe} \in {\BC} \ , 
\label{eq:qfug}
\eeq
$|{\qe} | < 1$. 

\end{enumerate}
We also use the notations: for any $a \in {\4}$, 
\beq
q_{a} ({\beta}) = e^{{\beta}{\ve}_{a}}, \quad P_{a} ({\beta}) = 1 - q_{a}({\beta}), \quad q_{a}^{*} ({\beta}) = e^{-{\beta}{\ve}_{a}}, \qquad P_{a}^{*} ({\beta}) = 1 - q_{a}^{*}({\beta}), 
\label{eq:qapa}
\eeq
and for any $S \subset {\4}$
\beq 
q_{S}({\beta}) = \prod_{a \in S} \, q_{a}({\beta}),  \quad q_{S}^{*} ({\beta}) = \prod_{a \in S} \, q_{a}^{*}({\beta}),  \quad P_{S}({\beta}) = \prod_{a \in S} \, P_{a}({\beta}), \quad P_{S}^{*}({\beta}) = \prod_{a \in S} \, P_{a}^{*}({\beta})
\label{eq:46not}
\eeq
We shall often skip the argument $\beta$ in the notations for $q_{a}, P_{S}^{*}$, etc. The notation \eqref{eq:46not}, in particular, implies (cf. \eqref{eq:epar}) 
\beq
q_{\4} =  q_{\emptyset} = 1, \quad P_{\4} = P_{1}P_{2} P_{3} P_{4} = P_{\4}^{*} \ , \quad q_{\bar A} = q_{A}^{*}
\eeq
and
\beq
P_{S}^{*} = (-1)^{|S|} q_{S}^{*} P_{S}
\eeq
We shall also encounter the relation
\beq
P_{\4} = P_{\3} + P_{\3}^{*}
\label{eq:4from3}
\eeq 
in what follows.

 \subsubsec{Geometry of the local model data}
 
 The meaning of the parameters $\ba$, $\bar\ve$ is the following. Define the gauge group $G_A$ corresponding to the stratum $X_A \approx {\BC}^{2}_{A}$ of the singular toric surface $X$ to be 
\beq
G_A = U( n_A )
 \label{eq:gagr}
 \eeq
 Let $T_A \subset G_A$ denote its maximal torus. 
 Let $U(1)^3_{\ve} \subset SU(4)$ be the maximal torus of the $(4,0)$-volume preserving unitary symmetries of $Z = {\BC}^{4}$. 
 The $U(1)^3_{\ve}$-action preserves $X$. 
 The Lie algebra ${\rm Lie}T_{A} \otimes {\BC}$ is parametrized by diagonal matrices
 ${\ba}_{A} = {\rm diag} \left( {\ac}_{A, 1} , \ldots , {\ac}_{A, n_{A}} \right)$ with complex entries ${\ac}_{A, {\alpha}} \in {\BC}$. The Lie algebra ${\rm Lie}U(1)^{3}_{\ve} \otimes {\BC}$ is parametrized by ${\rm diag}({\ve}_{1}, {\ve}_{2}, {\ve}_{3}, {\ve}_{4})$ with ${\ve}_{1} + {\ve}_{2} + {\ve}_{3} + {\ve}_{4} = 0$. 
 Let
\beq
T_{\Hf} = U(1)^{3}_{\ve} \, \times \,  \varprod_{A \in {\6}} T_{A} 
\eeq

\subsubsec{Additive to multiplicative}
Let $I_{\pm}$ be two finite sets, $I = I_{+} \amalg I_{-}$. 
Let ${\Mf}$ be a space with an action of a Lie group $\Gf$, and 
let ${\CalE}_{i}$, $i \in I$ be a collection of $\Gf$-equivariant vector bundles over
$\Mf$. Let $w_{i} \in {\BC}$.  We combine them into the ${\Gf} \times \left( {\BC}^{\times} \right)^{I}$-equivariant
virtual bundle ${\CalE} = \left[ \bigoplus_{i \in I_{+}} {\CalE}_{i} \, \ominus\,
\bigoplus_{i \in I_{-}} {\CalE}_{i} \right]$. Let 
\beq
{\rm Ch}_{\beta}({\CalE}_{i}) = \sum_{\alpha} e^{{\beta}{\xi}_{i,\alpha}}
\eeq
be the refined Chern character (with $\xi_{i,\alpha}$ equivariant Chern roots of $\CalE_{i}$),
so that in the non-equivariant setting 
\[ {\rm Ch}_{\beta}(E) = \sum_{k \geq 0} {\beta}^{k} {\rm ch}_{k}(E) \ , \]
and define 
\beq
f ({\beta}) = \sum_{i \in I_{+}} e^{{\beta}w_{i}} {\rm Ch}_{\beta}({\CalE}_{i}) -  \sum_{i \in I_{-}} e^{{\beta}w_{i}} {\rm Ch}_{\beta}({\CalE}_{i}) \, 
\label{eq:fb2}
\eeq
To $f({\beta})$ we associate the equivariant characteristic class, a rational function of $w_{i}$'s:
\beq
{\ep} [f] =  \prod_{i \in I_{+}} {\cb}_{w_{i}}({\CalE}_{+,i}) \, \prod_{i \in I_{-}} {\cb}_{w_{i}}({\CalE}_{i})^{-1}\  . 
\label{eq:epf2} 
\eeq
where
\beq
{\cb}_{w}({\CalE}) = \sum_{k=0}^{{\rm rk}{\CalE}} \
w^{k}\, c_{{\rm rk}{\CalE} - k} ({\CalE})
\eeq
is the usual $\Gf$-equivariant Chern polynomial of $\CalE$, evaluated at $w \in H^{\bullet}_{{\BC}^{\times}}(pt) = {\BC}$, equivalently, it is the top ${\BC}^{\times} \times {\Gf}$-equivariant Chern class of $\CalE$.  
We define the $*$-operation on the expressions $f({\beta})$: 
\beq
f^{*} ({\beta}) \equiv \sum_{i \in I_{+}} e^{-{\beta}w_{i}} {\rm Ch}_{\beta}({\CalE}_{i}^{*}) -  \sum_{i \in I_{-}} e^{-{\beta}w_{i}} {\rm Ch}_{\beta}({\CalE}_{i}^{*})  = f(-{\beta})\, 
\label{eq:fb3}
\eeq
This definition is consistent with the notations \eqref{eq:qapa}. 

We have:
\beq
{\ep} [ f ] = (-1)^{f(0)} {\ep} [ f^{*} ]
\label{eq:reflect}
\eeq
where
\beq
f(0) = \sum_{i \in I_{+}}  {\rm rk}({\CalE}_{i}) -  
\sum_{i \in I_{-}} {\rm rk}({\CalE}_{i}) \, 
\label{eq:fb4}
\eeq

Therefore, 
\beq
{\ep} [ P_{S} f ] = {\ep} [ P_{S} q_{\bar S} f^{*} ]^{(-1)^{|S|}}\, , 
\label{eq:refl2}
\eeq
The definition \eqref{eq:epf2} is the generalization of the notation used in 
\cite{Nekrasov:2016qym}, where 
we defined ${\ep}$ as a map from the space of ${\BZ}$-linear combinations of exponents to rational functions:
\beq
{\ep}\Biggl[  \, \sum_{i \in I_{+}} e^{{\beta}w_{i}} - \sum_{i \in I_{-}} e^{{\beta}w_{i}}\, \Biggr] \, = \, \left( \prod_{i \in I_{+}} w_{i} \right) \, \cdot \, \left( \prod_{i \in I_{-}} w_{i} \right)^{-1} \  . 
\label{eq:epf1} 
\eeq

\secc{Partition\ function\ of \ spiked\ instantons}

In this section we define the statistical mechanical model. The random variables are the strings of Young diagrams and the complex Boltzmann weights are rational functions of the 
complex numbers \eqref{eq:vep}, \eqref{eq:avec}. The definition might look first a bit artificial. Its origin is geometric. Namely, in \cite{Nekrasov:2016qym} the moduli space
of spiked instantons ${\iM}(k,{\vec n})$ is introduced, with ${\vec n} = (n_{A})_{A \in {\6}}$. It has an action of the group ${\Hf} = \times_{A \in \6} U(n_A) \times U(1)^{3}_{\ve}$. 
The fixed points of the maximal torus $T_{\Hf}$ are in one-to-one correspondence with the 
strings $\bla$ of partitions described below. The Boltzmann weight is simply the localization 
contribution to the integral of $1$ over ${\iM}(k,{\vec n})$, multiplied by ${\qe}^{k}$. 
This contribution is the product of the weights of the $T_{\Hf}$-action on the virtual tangent space to ${\iM}(k,{\vec n})$, which is the difference of the kernel and the cokernel of the linearization of the equations defining ${\iM}(k,{\vec n})$ at the fixed point. The kernel is always a complex vector space, henceforth it is naturally oriented and the product of weights is well-defined. The cokernel (the obstruction space) is only a real vector space, hence the product of the weights depends on the choice of its orientation. In what follows we specify the choice of the orientation with the help of the choice of the order on $\4$ and $\6$.  The resulting measure will not depend on this choice. 

\subsec{The configuration space}
The basic local model is a statistical ensemble. The random variables are the strings 
\beq
{\bla} = \left( {\lambda}^{(A, {\alpha})} \right)_{A \in {\6}, {\alpha} \in [n_{A}]} \in {\Lambda}^{{\bN}}
\label{eq:blabulk}
\eeq  of (cf. \eqref{eq:nlab})
$$
\sum_{A \in {\6}} n_{A} = {\#}{\bN}
$$
partitions ${\lambda}^{(A, \alpha)} \in {\Lambda}$.
In other words, the configuration space
is 
\beq
{\Lambda}^{{\bN}}
\, . \qquad 
\label{eq:confbulk}
\eeq
Define, cf. \eqref{eq:lamset}:
\beq
N_{A}({\beta}) = \sum_{{\alpha}=1}^{n_{A}} e^{{\beta}{\ac}_{A, \alpha}}
\, , \qquad 
K_{A} ({\beta}) = \sum_{{\alpha}=1}^{n_{A}} \sum_{{\square} \in {\lambda}^{(A, \alpha)}} e^{{\beta}c_{A, \alpha}({\square})} \, , \eeq
with
\beq
c_{A, \alpha}({\square}) = {\ac}_{A, \alpha} + {\ve}_{a} ( i - 1) + {\ve}_{b} (j-1), \qquad {\rm for} \ {\square} = (i,j) \label{eq:cont}
\eeq
and  (cf. \eqref{eq:fb3})
\beq
T_{A} \ = \ N_{A} K_{A}^{*} + \ q_{A}\, N_{A}^{*} K_{A} \ -\  P_{A}\, K_{A}K_{A}^{*}
\label{eq:tanchar}
\eeq
Let
\beq
k_{A} = K_{A}(0) = \sum_{{\alpha}=1}^{[n_{A}]} \ \vert \, {\lambda}^{(A, {\alpha})} \, \vert
\label{eq:instcha}
\eeq
and
\beq
| {\bla} | = \sum_{A \in \6} k_{A}
\eeq
It is well-known \cite{Nekrasov:2002qd, Nakajima:2003lectures, Alday:2009aq} that 
\beq
T_{A} = q_{A}T_{A}^{*}
\label{eq:sympl}
\eeq 
is a pure character, i.e. 
\beq
T_{A} = \sum_{I=1}^{2n_{A}k_{A}} e^{t_{A,I}}
\label{eq:tanchar2}
\eeq
where $t_{A,I}$ are integral linear combinations of ${\ac}_{A, \alpha}$, 
${\alpha} \in [n_{A}]$, 
${\ve}_{a}, {\ve}_{b}$, $a,b \in A$. Let us assume ${\ba}_{A}, {\bar\ve}$ are
sufficiently generic, so that $t_{A, I} \neq 0, t_{A,I} + {\ve}_{\bar a}  \neq 0$ for any ${\bar a} \in {\bar A}$, $I \in [2n_{A}k_{A}]$.

Define, finally,
\beq
K({\beta}) = \sum_{A \in \6} K_{A}({\beta})
\label{eq:ktot}
\eeq

\subsec{The statistical weight}

The complex Boltzmann weight of $\bla$ is given by the following expression:
\beq
Z_{\bla} = {\qe}^{|{\bla}|}\  {\ep} \left[ -T_{\bla} \right] \, , 
\label{eq:swbarlam}
\eeq
where (cf. \eqref{eq:gamap}):
\beq
T_{\bla} ({\beta})=  \sum_{A \in {\6}}   \left( P_{{\varphi}({\bar A})}T_{A} + 
P_{\bar A}  N_{A} \sum_{B\neq A} K_{B}^{*} \right) \, - \, P_{\4} \sum_{A < B} K_{A}K_{B}^{*}
\label{eq:tlam}
\eeq
The definition \eqref{eq:swbarlam} depends explicitly on the choice of the ordering of the sets $\4$ and $\6$, since it enters the definition of the maps ${\varphi}: {\6} \to {\4}$ and the meaning of $A < B$ in \eqref{eq:tlam}. Morally, 
\beq
 {\ep} \left[ -T_{\bla} \right] \sim {\ep} \left[ - \sum_{A \in \6} P_{\bar A}N_{A} K^{*} \right]
 \sqrt{{\ep} \left[ P_{\4} K K ^{*} \right]}
 \eeq
so the Boltzmann weight is defined canonically up to a sign. 

Note that for generic ${\ba}_{A}, {\bar\ve}$ the measure \eqref{eq:swbarlam}
does not depend on the choice of the order on $\4$ or $\6$:
\beq
\begin{aligned}
& {\ep} \left[ q_{\bar a} T_{A} \right] = {\ep} \left[ 
q_{\bar a}^{*} T_{A}^{*} \right] = {\ep} \left[ 
q_{\bar a}^{*} q_{A}^{*} T_{A} \right]  = {\ep} \left[ 
q_{\bar b} T_{A} \right] \, , \\
& {\ep} \left[ P_{\4} \sum_{A < B} K_{A}K_{B}^{*} \right] = {\ep} \left[ P_{\3} \sum_{A \neq B} K_{A}K_{B}^{*} \right] \\
\end{aligned}
\eeq
where we used \eqref{eq:sympl}, \eqref{eq:tanchar2}, \eqref{eq:4from3}, and $q_{\bar a} q_{\bar b}q_{A} = 1$ for ${\bar A} = \{ {\bar a}, {\bar b} \}$. 
Define, 
\beq
{\CalZ}^{\rm inst} = \sum\limits_{{\bla} \in {\Lambda}^{\bN}} \ Z_{\bla} \  = \ \sum_{k=0}^{\infty}\, {\qe}^{k} \, {\CalZ}^{\rm inst}_{k}
\label{eq:swzk}
\eeq
\subsubsection{The origins: spiked instantons, tori and characters}

The partition function ${\CalZ}^{\rm inst}$ is the $T_{\Hf}$-equivariant 
integral of $1$ over the virtual fundamental cycle of the moduli space
of spiked instantons \cite{Nekrasov:2016qym}. 
The latter is the space of solutions to certain quadratic matrix equations, generalizing the ADHM equations \cite{Atiyah:1978ri}, on four complex $k \times k$ matrices $B_{a}$, their Hermitian conjugates $B_{a}^{\dagger}$, $a \in \4$, and twelve rectangular matrices $I_{A}, J_{A}$, of sizes $n_{A} \times k$ and $k \times n_{A}$, $A \in \6$, and their Hermitian conjugates. 
The definition \eqref{eq:tlam} stems from the equivariant localization. 
The strings of partitions $\bla$ are the 
$T_{\Hf}$-fixed points. The matrices $(B_{a}, I_A, J_A)$ of the
construction \cite{Nekrasov:2016qym} obey, for such a fixed point:
\beq
[ B_{a}, B_{b} ] =0, \qquad a, b \in {\4}, \qquad J_A = 0, \qquad A \in {\6}
\eeq
the vectors
\beq
| i,j ; {\alpha} ; ab \rangle = 
B_{a}^{i-1}B_{b}^{j-1} I_{ab} (N_{ab, {\alpha}}) 
\eeq
with ${\alpha} \in [n_{ab}]$, $1 \leq j \leq {\lambda}^{(ab, {\alpha})}$
forming the basis of the vector space $K$, $N_{ab, {\alpha}}$
being the eigenspace of $T_{ab}$ action on the framing space 
$N_{ab}$ (see \cite{Nekrasov:2016qym} for the notations and more explanations). 
  
{}The equivariant weights of the matrices contribute
\beq
T_{+} = \sum_{a\in \4}  q_{a} K  K^{*} +  \sum_{A \in \6} \left( K^{*}N_{A} + q_{A} K N_{A}^{*}\right) 
\label{eq:matw}
\eeq
with
\beq
K = \sum_{A \in \6} K_{A}
\eeq
while the equivariant weights of the equations they obey, and the symmetries one divides by, contribute (with the minus sign)
\beq
T_{-} = \left( 1 + \sum_{c \in {\3}} q_{c}q_{4} \right)   K K^{*} + \sum_{A \in  \6} \, \sum_{{\bar a} \in {\bar A}} \, q_{\bar a}\, K^{*}N_{A}  
\label{eq:eqw}
\eeq
Moreover, the $T_{+}$ part is defined canonically by using the complex structure of the space of matrices $(B_{a}, I_{A}, J_{A})$. The $T_{-}$ part is defined non-canonically, as the expression \eqref{eq:eqw} does not respect the symmetry between $q_{a}$'s. The real (i.e. such that ${\chi}^{*}= {\chi}$) character $T_{-} + T_{-}^{*}$ is defined canonically. This subtlety has to do with the real, as opposed to complex, nature of the equations defining the spiked instantons  
\cite{Nekrasov:2016qym}. So, 
${\ep}[T_{-}]$ may have a sign ambiguity, as $\sqrt{{\ep}[T_{-} + T_{-}^{*}]}$. Also, ${\ep}[T_{-}]$ and ${\ep}[T_{+}]$ separately may vanish, as some of the weights in \eqref{eq:matw} and \eqref{eq:eqw} may vanish. It is easy to show that formally ${\ep}[T_{-} - T_{+}] = {\ep}[ -T_{\bla}]$. One simply uses \eqref{eq:reflect} several times.  
The details of the choice of the sign will be clarified elsewhere (it uses the residue definition of the localization contribution, which was worked out in \cite{Moore:1997dj}, it is similar to what sometimes is referred to as the Jeffrey-Kirwan residue in the mathematical literature). 

{}The resulting measure factor
\beq
{\ep}[T_{-} - T_{+}] = \frac{{\ep}[Obs_{\bla}]}{{\ep}[Def_{\bla}]} =  {\ep} [ - T_{\bla} ]
\eeq
where $Def_{\bla}$, $Obs_{\bla}$ are the $T_{\Hf}$-characters of  ${\rm ker}D_{\bla}$, ${\rm coker}D_{\bla}$, respectively. Here $D_{\bla}$ is the linearization of the spiked instanton equations at the solution, corresponding to $\bla$.

The expressions $N_A$, $K_A$, $T_{\bla} ({\beta})$ etc. are the elements of the K-group $K[T_{\Hf}]$, i.e. the abelian group whose elements are the formal linear combinations
\beq
\sum_{w \in T_{\Hf}^{\vee}} n_{w} L_{w}
\label{eq:kgrt}
\eeq
where $n_{w} \in {\BZ}$, 
\beq
L_{w}
\label{eq:irrept}
\eeq are the irreducible representations of the torus $T_{\Hf}$, i.e. the elements of the lattice $T_{\Hf}^{\vee} = {\rm Hom} \left( T_{\Hf}, U(1) \right)$. We assign to the weight $w = (w_{A, \alpha}) \oplus (w_{a})$ a function of $({\ba}, {\bar\ve})$, the character of $T_{\Hf}^{\BC}$ in the representation $L_w$:
\beq 
L_{w} \mapsto {\exp} \, {\beta} \left( \sum\limits_{A, \alpha} w_{A, \alpha}{\ac}_{A, \alpha} + 
\sum\limits_{a} w_{a} {\ve}_{a} \right)
\label{eq:chiw}
\eeq
Here $w_{A, \alpha} \in {\BZ}$, $w_{a} \in {\BZ}$ are defined up to a shift $w_{a} \mapsto w_{a} + {\sf w}$, ${\sf w} \in {\BZ}$. 

\subsubsec{More general definition}

The definition \eqref{eq:swzk} is fine as long as ${\ba}$ and ${\bar\ve}$ are generic.

However, e.g. if for some $ab \in {\6}$ the ratio ${\ve}_{a}/{\ve}_{b} \in {\bQ}_{+}$
is a positive rational number, or if for some ${\alpha} \neq {\beta} \in [n_A]$, ${\ac}_{A, {\alpha}} = {\ac}_{A, {\beta}}$, the individual contributions $Z_{\bla}$ to the formula \eqref{eq:swzk} have apparent poles. 
Actually, the poles cancel. Let us give the presentation of the formula \eqref{eq:swzk}
which is applicable in these cases. 
\beq
{\CalZ}^{\rm inst}_{k} \ = \ \sum_{(k_{A})_{A \in {\6}}, \ \sum_{A} k_{A} = k}\
\int_{\varprod_{A \in {\6}} {\iM}_{k_{A}}(n_{A})} \ {\CalS}_{\vec n, \vec k} ({\ba}, {\bar\ep})
\label{eq:spikedint}
\eeq
where Gieseker-Nakajima moduli  spaces ${\iM}_{k}(n)$ parametrize the charge $k$ noncommutative $U(n)$ instantons on ${\BR}^{4}$ 
and  framed rank $n$ torsion free sheaves $\mathcal{E}$
on ${\BC\BP}^{2}$ with ${\rm ch}_{2}({\mathcal{E}}) = k$, while ${\CalS}_{\vec n, \vec k} ({\ba}, {\bar\ep})$ is the equivariant characteristic class, given by (cf. \eqref{eq:tlam}):
\begin{multline}
{\CalS}_{\vec n, \vec k} ({\ba}, {\bar\ep}) \ = \ \prod_{A} {\cb}_{m_{A}}\left( T{\iM}_{k_{A}}(n_{A})\, \right) \ \times \qquad  \\
\\ \ \times\ \prod_{A \neq B \in {\6}} \prod_{{\alpha} \in [n_{A}]} \frac{\prod_{{\bar a} \in {\bar A}}   {\cb}_{{\ac}_{A, {\alpha}}+ {\ve}_{\bar a}} \left( K_{B} \right)}{ {\cb}_{{\ac}_{A, {\alpha}}} \left(  K_{B} \right) {{\cb}_{{\ac}_{A, {\alpha}}-{\ve}_{A}} \left( K_{B} \right)} }  \ \times \, \\ \qquad\qquad\qquad
 \prod_{A < B \in {\6}} \frac{\left( {\cb}_{0} \left( {\rm Hom}( K_{B} , K_{A} )\right) \right)^2  \prod_{C \in {\6}} {\cb}_{{\ve}_{C}}\left( {\rm Hom} ( K_{B} , K_{A} \right)}{\prod_{c \in {\4}} {\cb}_{{\ve}_{c}} \left( {\rm Hom} ( K_{B} , K_{A} ) \right) {\cb}_{-{\ve}_{c}} \left( {\rm Hom} ( K_{B} , K_{A} ) \right)}  \\
\end{multline}
with $K_{A}$ being the tautological rank $k_{A}$ bundle over ${\iM}_{k_{A}}(n_{A})$.
Finally, $m_{A} = {\ve}_{\bar a}$ for  either ${\bar a} \in {\bar A}$. The choice of $\bar a$ is immaterial. Indeed,  
the moduli space ${\iM}_{k_{A}}(n_{A})$ is a complex symplectic manifold, as reflected by the symmetry $T_{A} = q_{A} T_{A}^{*}$. It implies
\beq
{\cb}_{{\ve}_{\bar a}} \left( T{\iM}_{k_{A}}(n_{A}) \right) = {\cb}_{{\ve}_{\bar b}} \left( T{\iM}_{k_{A}}(n_{A}) \right)\eeq
for both ${\bar a}, {\bar b} \in {\bar A}$.

\subsubsec{One-instanton case}

As an illustration, let us consider the case $k=1$. There are 
\[ \sum_{A} n_{A} \]
possibilities, with
\beq
K_{B} = {\delta}_{A, B} \, e^{{\beta}{\ac}_{A, \alpha}}, \qquad A \in {\6}, \, {\alpha} \in [ n_{A} ] 
\label{eq:kb}
\eeq
Thus, the $1$-instanton partition function is given by:
\beq
{\CalZ}^{\rm inst}_{1}  \ =\  \sum_{A \in {\6}} \ \sum_{{\alpha} \in [ n_{A} ]}\, \ Z_{A, \alpha} \, ,  
\label{eq:z1aa}
\eeq
with
\beq
\begin{aligned}
Z_{A, \alpha} \ & =\   \frac{E_{\bar A} - E_{A}}{E_{A}}  \prod_{{\alpha}' \in [ n_{A}], \ {\alpha}' \neq {\alpha}} \, \left(  1 + 
\frac{E_{\bar A}}{\left( {\ac}_{A, {\alpha}'} - {\ac}_{A, \alpha}  \right)\left( {\ac}_{A, {\alpha}'} - {\ac}_{A, \alpha} - {\ve}_{A} \right)} \right) \ \times \\
& \qquad\qquad\qquad\qquad\qquad\qquad \times \ \prod_{B \neq A} \prod_{{\gamma} \in [n_{B}]}  \, \left( 1 + 
\frac{E_{\bar B}}{\left( {\ac}_{B, \gamma} - {\ac}_{A, \alpha}  \right) \left( {\ac}_{B, \gamma} - {\ac}_{A, \alpha} - {\ve}_{B} \right)} \right)
\end{aligned}
\eeq
where 
\beq
{\ve}_{S} = \sum_{s \in S} {\ve}_{s} = - {\ve}_{\bar S}, \qquad E_{S} = \prod_{s \in S} {\ve}_{s}
\eeq 

\subsec{The perturbative prefactor}

We introduce a common, i.e. $\bla$-independent
 prefactor in the statistical weight. The so completed statistical weight  is equal to:
\beq
W_{\bla} = Z^{\rm pert} ({\bar\ac}, {\bar\ve}) \, Z_{\bla} 
\label{eq:wlam} 
 \eeq
 with
\begin{multline}
Z^{\rm pert} ({\bar\ac}, {\bar\ve})  =  \prod\limits_{A \in {\6}} \ Z_{N=2*}^{{\rm pert}, A} ( {\bf\ac}_{A}, {\bar\ve} ) \qquad \times \\ 
 \prod\limits_{\{a , b, c \} \subset {\4}}\, Z^{{\rm pert}, a|bc}_{\rm fold} 
 ({\bf\ac}_{ab}, {\bf\ac}_{ac}, {\bar\ve} ) \qquad \times \\
  \prod\limits_{A \in {\6}, A< {\bar A}} \, Z^{{\rm pert}, A}_{\rm cross}
( {\bf\ac}_{A}, {\bf\ac}_{\bar A} , {\bar\ve}) \end{multline}
where 

$\bullet$ for $A = ab$: define $m_{A} = {\ve}_{{\varphi}(A)}$, and 
\beq
\begin{aligned}
& Z_{N=2*}^{{\rm pert}, A} ( {\bf\ac}_{A}, {\bar\ve} ) = 
\prod_{{\alpha}, {\beta} =1}^{n_{A}} {\Gamma}_{2} \left( 
{\ac}_{A, \alpha} - {\ac}_{A, \beta} ; {\ve}_{a}, {\ve}_{b} \right) \ \times \\
& \qquad\qquad\qquad  \prod_{{\alpha}, {\beta} =1}^{n_{A}} {\Gamma}_{2}^{-1} \left( 
{\ac}_{A, \alpha} - {\ac}_{A, \beta} + m_{A}; {\ve}_{a}, {\ve}_{b} \right)
\label{eq:zn2*}\end{aligned}
\eeq
with the Barnes double gamma functions ${\Gamma}_{2}(x; {\ve}_{a}, {\ve}_{b})$
normalized in such a way so as to have simple zeroes on a quadrant of the integral lattice spanned by ${\ve}_{a}, {\ve}_{b}$:
\beq
{\Gamma}_{2}(x; {\ve}_{a}, {\ve}_{b}) \sim \prod_{i, j \geq 1} \left( x + {\ve}_{a} (i-1) +{\ve}_{b}(j-1) \right)   \qquad \ , 
\eeq
it is defined by the analytic continuation of the integral formula
\beq
{\Gamma}_{2}(x; {\ve}_{a}, {\ve}_{b})  = {\exp} \, - \, \frac{d}{ds} \, \Biggr\vert_{s=0} \,\frac{1}{{\Gamma}(s)} \int_{0}^{\infty} \frac{dt}{t} t^{s} \, \frac{e^{-tx}}{(1-e^{-t{\ve}_{a}})(1-e^{-t{\ve}_{b}})}
\eeq
from the domain ${\rm Re}(x), {\rm Re}({\ve}_{a}), {\rm Re}({\ve}_{b}) > 0$.

$\bullet$
\beq
Z^{{\rm pert}, a|bc}_{\rm fold} ({\bf\ac}_{ab}, {\bf\ac}_{ac}, {\bar\ve} ) =  \ \prod_{{\alpha}=1}^{n_{ab}}
\prod_{{\beta}=1}^{n_{ac}}\
{\Gamma}_{1} \left( {\ac}_{ab, \alpha} - {\ac}_{ac, \beta} + {\ve}_{a} + {\ve}_{c}; {\ve}_{a} \right)  
\eeq
where ${\Gamma}_{1}(x; {\ve}_{a})$ is essentially the ordinary $\Gamma$-function:
\beq
 {\Gamma}_{1}(x; y)  \sim \prod_{i=1}^{\infty} \left( x + y(i-1) \right) \, , 
\eeq
Again, it can be defined by the analytic continuation of the integral
\beq
{\Gamma}_{1}(x; y)  = {\exp} \, - \, \frac{d}{ds} \, \Biggr\vert_{s=0} \, \frac{1}{{\Gamma}(s)} \int_{0}^{\infty} \frac{dt}{t} t^{s} \, \frac{e^{-tx}}{(1-e^{-ty})}
\eeq
from the domain ${\rm Re}(x), {\rm Re}(y) > 0$, giving
\beq
{\Gamma}_{1}(x; y)  =  \frac{\sqrt{2\pi / y}}{y^{\frac{x}{y}}\, {\Gamma}\left( \frac{x}{y} \right)} \, \ , 
\label{eq:gamma1}
\eeq

$\bullet$
\beq
Z^{{\rm pert}, A}_{\rm cross}
( {\bf\ac}_{A}, {\bf\ac}_{\bar A} , {\bar\ve}) \, = \, \prod_{{\alpha}=1}^{n_{A}} \prod_{{\beta}= 1}^{n_{\bar A}} \ \left( {\ac}_{A, \alpha} - {\ac}_{{\bar A}, \beta} +
 {\ve}_{\bar A} \right) \eeq
 
\subsubsection{Anomalies and other definitions of perturbative factors} In \cite{Alday:2009aq} another normalization for the perturbative prefactor is used: the second line in \eqref{eq:zn2*} would read, in our notation, as
$$
\prod_{1 \leq {\alpha} < {\beta} \leq n_{A}} \ \prod_{{\bar a} \in {\bar A}} 
\ {\Gamma}_{2}^{-1} \left( {\ac}_{A, \alpha} - {\ac}_{A, \beta} + {\ve}_{\bar a}\, ; \, {\ve}_{a}, {\ve}_{b} \right) 
$$
This normalization makes explicit the symmetry between ${\bar a} \in {\bar A}$, however the gauge invariance, i.e. the Weyl symmetry of $U(n_A)$ acting on ${\ac}_{A, {\alpha}}$ is partly broken.  

{}Unlike the instanton partition function ${\CalZ}^{\rm inst}$ the perturbative
factor does depend on the choice of the order on $\4$ which is used in the definition of
$m_{A} = {\ve}_{{\varphi}(A)}$. This dependence will be analyzed elsewhere. 

\subsubsection{Subtleties for tuned parameters}
 
When the equivariant parameters ${\ba}, {\bar\ve}$ are rationally dependent, the torus ${\tilde T}_{\Hf} \subset T_{\Hf}$ they generate is strictly smaller than $T_{\Hf}$.  Accordingly, the fixed points on the moduli space of spiked instantons need not be isolated, and the formula \eqref{eq:spikedint} is used. It can be further localized to the
set of torus-fixed points on ${\iM}_{k_{A}}(n_{A})$, which are relatively well-understood
in the case $n_{A}=1$ \cite{iarrobino:1977, Iarrobino:1997, Loginov:2014}. We shall encounter these complications when dealing with gauge theories on ${\BC}^{2}/{\Gamma}$ spaces, or on the complex surfaces in the ${\BC}^{4}/{\Gamma}^{'} \times {\Gamma}^{''}$
spaces, with finite $SU(2)$ subgroups $\Gamma$, $\Gamma'$, $\Gamma'' \subset SU(2)$ of $D$ or $E$ type. 
  
 \subsec{The main result}
 
Here is the main fact about the partition function of spiked instantons:  
the compactness theorem proven in  \cite{Nekrasov:2016qym}
implies  
${\CalZ}_{\rm spiked} ({\ba}, {\bar \ve})$
defined by
\beq
{\CalZ}_{\rm spiked} ({\ba}, {\bar \ve}) \ = \ \sum_{\bla} \, W_{\bla}
\eeq
has no singularities in the variables:
\beq
x_{A} \ = \ \frac{1}{n_{A}} \sum_{{\alpha}=1}^{n_{A}} \, {\ac}_{A, {\alpha}}
\label{eq:centofmass}
\eeq
with fixed 
\beq
{\tilde\ac}_{A, {\alpha}} \ = \  {\ac}_{A, {\alpha}} - x_{A}
\label{eq:tildaca}
\eeq
$\underline{Remark}$. The reason we have to keep the majority of our variables
fixed is the denominator ${\Gamma}_{2}^{-1}$ in the perturbative prefactors $Z_{N=2*}^{\rm pert, A}$. Without it the partition function would have been an entire function of all ${\ac}_{A, \alpha}$'s.

\secc{Orbi\  folding}

In this section we discuss the partition function of the generalized gauge theories defined on the orbifolds with respect to a discrete (finite) group ${\Gammadi}$. Both the worldvolume and the transverse space of the theory may be subject to the orbifold projection. 
Geometrically, the action of $\Gammadi$ factors through the linear action in ${\BC}^{4}$, which
we assume to preserve the Calabi-Yau fourfold structure:
\beq
{\bro}_{\rm geom}: {\Gammadi} \longrightarrow SU(4)
\label{eq:rhog}
\eeq
This construction is motivated by the consideration of D-branes on ${\BC}^{4}/{\Gammadi}$. As is explained in \cite{Douglas:1996sw}, the orbifold projection involves an action of $\Gammadi$ on the Chan-Paton spaces:
\beq
{\bro}_{\rm CP}: {\Gammadi} \longrightarrow \varprod_{A \in {\6}} U(n_{A})
\label{eq:cprep}
\eeq 
which amounts to the decomposition:
\beq
N_{A} = \bigoplus_{{\bom} \in {\Gammadi}^{\vee}} N_{A, {\bom}} \otimes {\CalR}_{\bom}
\label{eq:deccp}
\eeq
The global symmetry group $\Hf$ is reduced to the $\Gammadi$-centralizer: the subgroup ${\Hf}_{\Gammadi} \subset {\Hf}$ which commutes with $\Gammadi$. 
The particular cases of this construction are the quiver gauge theories, the theories in the presence of special surface operators, possibly intersecting, and the theories on the ALE spaces. 

The parameters of the partition function of the orbifolded theory are 
$({\widetilde\ba}, {\widetilde{\ve}}) \in {\rm Lie}T_{{\Hf}_{\Gammadi}} \otimes {\BC}$, where ${\widetilde\ba}$ is in the Cartan subalgebra of the centralizer of the image ${\bro}_{\rm CP}({\Gammadi})$ in \eqref{eq:cprep}, while $\widetilde{\ve}$ is in the Cartan subalgebra of the centralizer of the image of ${\bro}_{\rm geom}({\Gammadi})$ in $SU(4)$. In addition, the fugacity $\qe$ of the original model fractionalizes:
\beq
{\qe} \longrightarrow {\bqe} = \left( {\qe}_{\bom} \right)_{{\bom} \in {\Gammadi}}
\eeq

\subsubsection{Choices of discrete groups}

Since we want the action of $\Gammadi$ to admit the invariant complex two-planes  supporting the strata $(X_{A}, G_{A})$ of the generalized gauge theories, at least for one $A \in {\6}$, the choice of $\Gamma$ is reduced to the following three possibilities:
\begin{enumerate}

\item The {\bf abelian} case:
\beq
{\Gammadi} = {\Gamab}, \eeq
represented in $U(1)_{\ve}^{3} \subset SU(4)$ with the help of the homomorphism ${\bro}_{\rm geom} = {\rm diag} ({\bro}_{a} )_{a\in \4}$:
\beq
{\bro}_{a} \left[ \ \left( \ e^{ \, \frac{2\pi \sqrt{-1}m_{\kappa}}{p_{\kappa}^{l_{\kappa}}} \, } \ \right)_{\kappa} \, \right] = {\exp}\, \left( 2\pi\sqrt{-1} \sum_{\kappa} \frac{m_{\kappa} n^{\kappa}_{a}}{p_{\kappa}^{l_{\kappa}}} \right)
\label{eq:rhoa}
\eeq
where $n^{\kappa}_{a} = 0, \ldots , p_{\kappa}^{l_{\kappa}}-1$, and
\beq
\sum_{a \in \4} n_{a}^{\kappa} = p_{\kappa}^{l_{\kappa}} n^{\kappa}\, \qquad n^{\kappa} \in {\BZ}\, , \label{eq:nak}
\eeq
The Chan-Paton representation \eqref{eq:cprep} amounts to the choice  of multiplicities $n_{A, \bnu}$:
\beq
{\bro}_{\rm CP}: \left(  e^{ \,  2\pi \sqrt{-1}m_{\kappa}p_{\kappa}^{-l_{\kappa}} \, } \right)_{\kappa} \mapsto 
{\rm diag} \left( \ \prod_{\kappa} e^{\,  2\pi \sqrt{-1} m_{\kappa}
 n^{\kappa}p_{\kappa}^{-l_{\kappa}}} \,  \cdot {\bf 1}_{n_{A, \bnu}} \ \right)_{{\bnu} \in {\Gamav}} \in\ U(n_{A})
 \eeq
 where ${\bnu} = \left( n^{\kappa} \right)_{\kappa} \in {\Gamav}$ labels the irreducible (one-dimensional) representations of $\Gamab$. 
The centralizer ${\Hf}_{\Gammadi}$ is equal to
\beq
{\Hf}_{\Gammadi} = U(1)^{3}_{\ve} \ \times \ \varprod_{{\bnu} \in {\Gamav}} \, \varprod_{A \in {\6}}   \ U(n_{A, {\bnu}}) \, , 
\eeq
its maximal torus
\beq
T_{{\Hf}_{\Gammadi}} = U(1)^{3}_{\ve} \ \times \ \varprod_{{\bnu} \in {\Gamav}} \, \varprod_{A \in {\6}}   \ T_{n_{A, {\bnu}}} \, , 
\eeq
with $T_{n_{A,{\bnu}}} \subset U(n_{A,{\bnu}})$ the maximal torus of diagonal matrices. 

Define (cf. \eqref{eq:nlab}):
\beq
{\bN}_{\Gamab} = \bigsqcup_{A\in {\6}, {\bnu} \in {\Gamav}}\ [n_{A, {\bnu}}] = \{ \, (A, {\bnu}, {\alpha}) \, | \, A \in {\6}, \, {\bnu} \in {\Gamav}, \, {\alpha} \in [n_{A, {\bnu}}] \, \}
\label{eq:nlabab}
\eeq

\bigskip

\item The  {\bf abelian} $\times$  {\bf ALE} case:
\beq
{\Gamma} = {\Gamab} \times {\Gamma}_{\gamma} \label{eq:nonab12}
\eeq
represented in $SU(4)$
with the help of the homomorphism 
\beq
{\bro}_{\rm geom} = \left(  \begin{matrix} & \begin{matrix} & {\bro}_{1}= {\bro}_{l}{\bro}_{r} & \\
& &  {\bro}_{2} = {\bro}_{l}{\bro}_{r}^{-1} \end{matrix}  &   \begin{matrix} &  & \\
& &   \end{matrix} \\
&  \begin{matrix} &  & \\
& &   \end{matrix} &  {\bro}_{34} = {\bro}_{l}^{-1} T_{\bf 2} \in U(2)_{34} \end{matrix}  \right) \  , 
\eeq
with
\beq
{\bro}_{\alpha} \left[ \ \left( e^{2\pi \sqrt{-1}m_{\kappa}p_{\kappa}^{-l_{\kappa}}}  \right)_{\kappa} \, \times \, h \ \right] \ = \ \prod_{\kappa} \, e^{\, 2\pi\sqrt{-1}m_{\kappa}  {\rho}^{\kappa}_{\alpha}p_{\kappa}^{-l_{\kappa}}} \qquad \in\ U(1) \, , \ {\alpha} = l, r  
\eeq
where ${\rho}^{\kappa}_{\alpha} = 0, \ldots , p_{\kappa}^{l_{\kappa}}-1$, and
\beq
{\bro}_{34} \left[ \ \left( e^{2\pi \sqrt{-1}m_{\kappa}p_{\kappa}^{-l_{\kappa}}} \right)_{\kappa} \, \times \, h \ \right] \ = \ \prod_{\kappa} \, e^{\, - 2\pi\sqrt{-1}m_{\kappa}  {\rho}^{\kappa}_{l}p_{\kappa}^{-l_{\kappa}}} T_{\bf 2}(h) 
\label{eq:34hom}
\eeq
with $T_{\bf 2}$ the defining two-dimensional representation of $SU(2) \ni {\Gamma}_{\gamma}$. 

The irreducible representations of the group $\Gammadi = {\Gamab} \times {\Gamma}_{\gamma}$ are the tensor products:
\beq
{\CalR}_{\bom} = {\CalL}_{\bnu} \otimes R_{\ib}  \, , 
\eeq
 labelled by the pairs ${\bom} = ({\bnu}, {\ib})$, ${\bnu} \in {\Gamav}$, ${\ib} \in {\Vg}$.
With the choice \eqref{eq:nonab12} of $\Gamma$ the only non-trivial Chan-Paton spaces are $N_{12}$ and $N_{34}$. The choice of the Chan-Paton representation ${\bro}_{\rm CP}$  in this case amounts to the choice of multiplicity spaces $N_{12,   \bom}, N_{34,   \bom}$, i.e. the dimension vectors
\beq
n_{\bom} = {\rm dim} N_{12, \bom}\, , \qquad
w_{\bom} = {\rm dim} N_{34,  \bom} \, , 
\eeq
The centralizer 
\beq
{\Hf}_{\Gammadi} = U(1)^{2}_{\gamma} \ \times \ \varprod_{{\bom} \in {\Gammadi}^{\vee}} U(n_{\bom}) \times  U(w_{\bom})
\label{eq:centra1234}
\eeq
where $U(1)^2_{\gamma} \subset U(1)^{3}_{\ve}$ consists of the diagonal matrices of the form:
\beq
 \left(  \begin{matrix} & \begin{matrix} & e^{\sqrt{-1}{\vartheta}_{1}} & \\
& &  e^{\sqrt{-1}{\vartheta}_{2}} \end{matrix}  &   \begin{matrix} &  & \\
& &   \end{matrix} \\
&  \begin{matrix} &  & \\
& &   \end{matrix} &  e^{-\frac{\sqrt{-1}}{2} \left( {\vartheta}_{1} + {\vartheta}_{2} \right)} \cdot {\bf 1}_{\bf 2} \end{matrix}  \right) \  \in SU(4)  , 
\eeq 

Define (cf. \eqref{eq:nlab}, \eqref{eq:nlabab}), for ${\Gammadi}= {\Gamab} \times {\Gamma}_{\gamma}$:
\beq
{\bN}_{\Gammadi} = {\bN}_{\Gammadi}^{+} \ \sqcup {\bN}_{\Gammadi}^{-}
\label{eq:nlababale}
\eeq
with
\begin{multline}
 {\bN}_{\Gammadi}^{+} =  \bigsqcup_{{\bom} \in {\Gammadi}^{\vee}}\ [n_{\bom}] = \left\{ \, ({\bom}, {\alpha}) \, | \,  {\bom} \in {\Gammadi}^{\vee}, \, {\alpha} \in [n_{{\bom}}] \, \right\}  \, , \qquad 
 {\bN}_{\Gammadi}^{-} =  \bigsqcup\limits_{\ib \in \Vg} \, {\bN}_{\Gammadi}^{{\ib}, -} \, , \\
  {\bN}_{\Gammadi}^{{\ib}, -} =  \bigsqcup_{{\bnu} \in {\Gamav}}\ [w_{{\ib}, \bnu}] = \left\{ \, ({\ib}, {\bnu}, {\beta}) \, | \,  {\bnu} \in {\Gamav}, \, {\beta} \in [w_{{\ib}, {\bnu}}] \, \right\} \\
\label{eq:nlababale2}
\end{multline}

\item The {\bf ALE $\times$ ALE} case:
\beq
{\Gammadi} = {\Gamab} \times {\Gamma}_{{\gamma}'} \times {\Gamma}_{{\gamma}''}
\label{eq:gamma1234}
\eeq
represented in $SU(4)$
with the help of the homomorphism 
$$
{\bro}_{\rm geom} \left[ t \times h' \times h''  \right]  = \left(  \begin{matrix} &{\bro}(t) \cdot T_{\bf 2} (h')  & 0 \\
& 0 &  {\bro}(t)^{-1} \cdot T_{\bf 2} (h'') \end{matrix} \ \right) \in SU(4)  , 
$$ 
with $h' \in {\Gamma}_{{\gamma}'}$, $h'' \in {\Gamma}_{{\gamma}''}$,  ${\bro}  \in {\Gamav}$. The irreducible representations of $\Gammadi$ are the tensor products 
\beq
{\CalR}_{\bom} = {\CalL}_{\bnu} \otimes R'_{\ib'} \otimes R''_{\ib''}
\eeq 
labelled by ${\bom} = ({\bnu}, {\ib}', {\ib}'')$, where ${\bom}  \in {\Gamav}$, ${\ib}' \in {\rm Vert}_{\gamma'}$, ${\ib}'' \in {\rm Vert}_{\gamma''}$, and $R'$, $R''$ are the irreps of ${\Gamma}_{\gamma'}$, ${\Gamma}_{\gamma''}$, respectively. Again, with the choice \eqref{eq:gamma1234} of $\mathbf\Gamma$ the only non-trivial Chan-Paton spaces are $N_{12}$ and $N_{34}$. The choice of the Chan-Paton representation ${\bro}_{\rm CP}$  in this case amounts to the choice of multiplicity spaces $N_{12, \bom}, N_{34,  \bom}$, i.e. the dimension vectors
\beq
n_{\bom} = {\rm dim} N_{12, \bom}\, , \qquad
w_{\bom} = {\rm dim} N_{34,  \bom} \, , 
\eeq
for ${\bom} = \left( {\bnu}, \ib', {\ib}'' \right) \in {\Gammadi}^{\vee}$. 
 The centralizer 
 \beq
{\Hf}_{\Gammadi} = U(1)_{\gamma' , \gamma''} \ \times \ \varprod_{{\bom} \in {\Gammadi}^{\vee}} U(n_{\bom}) \times  U(w_{\bom})
\label{eq:centra1234}
\eeq
where $U(1)_{\gamma', \gamma''} \subset U(1)^{3}_{\ve}$ consists of diagonal matrices of the form:
\beq
 \left(  \begin{matrix} &  e^{\sqrt{-1}{\vartheta}} \cdot {\bf 1}_{\bf 2} & \\
&   &  e^{- \sqrt{-1} {\vartheta}} \cdot {\bf 1}_{\bf 2} \end{matrix}  \ \right) \  \in SU(4)  , 
\eeq 
\end{enumerate}
Define (cf. \eqref{eq:nlab}, \eqref{eq:nlabab}, \eqref{eq:nlababale}), for ${\Gammadi}= {\Gamab} \times {\Gamma}_{\gamma'} \times {\Gamma}_{\gamma''}$:
\beq
{\bN}_{\Gammadi} = {\bN}_{\Gammadi}^{+} \ \sqcup \ {\bN}_{\Gammadi}^{-}
\label{eq:nlababaleale}
\eeq
with ${\bN}_{\Gammadi}^{+} = \bigsqcup\limits_{\ib' \in {\rm Vert}_{\gamma'}} \, {\bN}_{\Gammadi}^{{\ib}', +}$, ${\bN}_{\Gammadi}^{-} =  
 \bigsqcup\limits_{\ib'' \in {\rm Vert}_{\gamma''}} \, {\bN}_{\Gammadi}^{{\ib}'', -}$, and
\begin{multline}
 {\bN}_{\Gammadi}^{{\ib}', +}  \ = \  \bigsqcup_{{\bnu} \in {\Gamav}, \, {\ib}'' \in {\rm Vert}_{\gamma''}}\ [n_{{\bnu}, {\ib}', {\ib''}}] = \left\{ \, ({\bnu}, {\ib}', {\ib}'', {\alpha}) \ | \  {\bnu} \in {\Gamav}, \, {\ib}'' \in {\rm Vert}_{\gamma''}, \, {\alpha} \in [n_{{\bnu}, {\ib}', {\ib''}}] \, \right\} \, , \\
 {\bN}_{\Gammadi}^{{\ib}'', -} \ = \  \bigsqcup_{{\bnu} \in {\Gamav}, \, {\ib}' \in {\rm Vert}_{\gamma'}}\ [w_{{\bnu}, {\ib}', {\ib''}}] = \left\{ \, ({\bnu}, {\ib}', {\ib''}, {\beta}) \ | \  {\bnu} \in {\Gamav}, \, {\ib}' \in {\rm Vert}_{\gamma'}, \, {\beta} \in [w_{{\bnu}, {\ib}', {\ib''}}] \, \right\} \\
\label{eq:nlabaleale2}
\end{multline}

In what follows the expressions $N_A$, $K_A$ etc. are promoted to ${\CalN}_{A}$, ${\CalK}_{A}$ etc. which are valued in $K[T_{{\Hf}_{\bf\Gamma}}] \otimes K[{\bf\Gamma}]$, 
i.e. they are the formal linear combinations:
\beq
\sum_{w \in T_{{\Hf}_{\Gammadi}}^{\vee}\, , \ {\bom} \in {\Gammadi}^{\vee}} \ n_{w, {\bom}} \ L_{w} \otimes {\CalR}_{\bom}
\eeq
where $L_w$ are the characters of $T_{{\Hf}_{\Gammadi}}$, and ${\CalR}_{\bom}$
are the irreducible representations of $\Gammadi$. Likewise, the ``tangent space'' character $T_{A}$ is promoted to ${\CalT}_{A} \in K[T_{{\Hf}_{\Gammadi}}] \otimes K[{\Gammadi}]$.

\subsubsection{Orbifold partition functions}

The definition of the partition function in the orbifold situation is the following. The random variable is a string $\bla$ of objects, which now involve both Young diagrams and connected components of the Nakajima-Young varieties, specifically:

\begin{enumerate}

\item{}

In the {\bf abelian} case the random variables are again the Young diagrams (partitions) 
${\lambda}^{(A, {\bnu}, {\alpha})} \in {\Lambda}$, now labeled by triples: :
\beq
{\bla}_{\rm ab} = \left( \, {\lambda}^{(A, {\bnu}, {\alpha})} \, \right)_{A \in {\6}, \, {\bnu} \in {\Gamav}, \, {\alpha} \in [n_{A, {\bnu}}]} \in {\Lambda}^{{\bN}_{\Gamab}}
\label{eq:strabobj}
\eeq

\item{}

In the {\bf abelian} $\times$  {\bf ALE}  case the random variables 
are the collections of of two types of objects: Young diagrams as before, and the connected components of Nakajima-Young varieties:
\begin{multline}
{\bla}_{{\rm ab} \times {\rm ale}} = \left( \ \left( \,  {\lambda}^{({\bom}, {\alpha})} \, \right)_{{\bom} \in {\Gammadi}^{\vee}, \, {\alpha} \in [n_{\bom}]} \, ; \,  \left( {\mu}^{({\bom}, {\beta})} \right)_{{\bom} \in {\Gammadi}^{\vee}, \, {\beta} \in [w_{\bom}]}\ \right) \\
\in {\Lambda}^{{\bN}^{+}_{\Gammadi}} \ \times \, \varprod\limits_{\ib \in \Vg} \, \left( {\Lambda}^{\ib}_{\gamma} \right)^{{\bN}^{{\ib}, -}_{\Gammadi}}
\label{eq:abalec}
\end{multline}
with ${\lambda}^{({\bom}, {\alpha})} \in \Lambda$, ${\bom} \in {\Gammadi}^{\vee}$, ${\mu}^{(({\ib}, {\bnu}), {\beta})} \in {\Lambda}^{\ib}_{\gamma}$, for ${\ib} \in \Vg$, ${\bnu} \in {\Gamav}$.

\item{}

Finally, in the
{\bf ALE} $\times$ {\bf ALE} case the random variables 
are the collections of  connected components of Nakajima-Young varieties:
\beq
{\bla}_{{\rm ale} \times {\rm ale}} = \left( \ \left( \,  {\mu}^{({\bom}, {\alpha})} \, \right)_{{\bom} \in {\Gammadi}^{\vee}, \, {\alpha} \in [n_{\bom}]} \, ; \,  \left( {\tilde\mu}^{({\bom}, {\beta})} \right)_{{\bom} \in {\Gammadi}^{\vee}, \, {\beta} \in [w_{\bom}]}\ \right)
\label{eq:alealec}
\eeq
with ${\mu}^{({\bom}, {\alpha})} \in {\Lambda}^{\ib'}_{\gamma'}({\bv'})$, ${\tilde\mu}^{({\bom}, {\beta})} \in {\Lambda}^{\ib''}_{\gamma''}({\bv''})$, with ${\bom} = ({\bnu}, {\ib}', {\ib}'')$, ${\bnu} \in {\Gamav}$, ${\ib}' \in {\rm Vert}_{\gamma'}$, ${\ib}'' \in {\rm Vert}_{\gamma''}$.

\end{enumerate}

We first describe the case of abelian orbifolds, and then proceed with the somewhat more restricted case of the non-abelian orbifolds. In the latter case our formulas are less explicit. 

\subsec{Abelian orbifolds}

We define the statistical model, which is parametrized by the following generalization of the data of the section $\bf 2$: 

\begin{enumerate}
\item{} The string $\bar\ve =  ({\ve}_{a})_{a \in \4} $ of $4$ complex numbers ${\ve}_{a}$, $a \in {\4}$ which sum up to zero, as in \eqref{eq:epar}

\item{} The string ${\bro}_{\rm geom} = ({\bro}_{a})_{a \in \4}$ of $4$ irreducible $\Gamab$-representations ${\bro}_{a}  \in {\Gamav}$ obeying
\beq
\sum_{a \in {\4}} {\bro}_{a} = 0 \in {\Gamav}
\eeq 
In other words, ${\bro}$ is a homomorphism ${\Gamab} \to U(1)^{3}_{\ve} \subset SU(4)$, so that
\beq
{\square}_{SU(4)} = \bigoplus_{a \in {\4}} \, {\CalL}_{{\bro}_{a}}
\eeq 
We shall also use the notation $\rho_{S}$ with $S \subset {\4}$ for the sum:
\beq
{\bro}_{S} = \sum_{s \in S} {\bro}_{s} \ . 
\label{eq:tenss}
\eeq
so that ${\bro}_{\emptyset} = {\bro}_{\4} = {\bro}_{0}$ and
\beq
{\wedge}^{\bullet}{\square}_{SU(4)} = \bigoplus_{S \subset {\4}} {\CalL}_{{\bro}_{S}} 
\eeq

\item{} The string $\bar n$ of $6$ $\Gamab$-representations 
\beq
N_{A} = \bigoplus\limits_{{\bnu} \in {\Gamav}} N_{A, \bnu} \otimes {\CalL}_{\bnu}, \qquad A \in {\6}
\label{eq:narep}
\eeq
with the multiplicity spaces $N_{A, \bnu} \approx {\BC}^{n_{A, \bnu}}$ of dimensions $n_{A, \bnu} = {\rm dim}N_{A, \bnu}$. 

\item{} The string $\bar\ac = \left({\ac}_{A, \bnu, \alpha}\right)_{{\alpha} \in [n_{A, {\bnu}}], \bnu \in {\Gamav}} \in {\BC}^{{\bN}_{\Gamab}}$ of   
\beq
\sum_{A \in {\6}, {\bnu} \in {\Gamav}} n_{A, \bnu} = {\#}{\bN}_{\Gamab}
\eeq
complex numbers ${\ac}_{A, {\bnu}, \alpha}$, ${\alpha} \in [n_{A, \bnu}]$. 

The data $\left( {\bar\ac}; {\bar\ve} \right)$ parametrizes the Cartan subalgebra of the centralizer ${\Hf}_{\Gamab}$. Define,  for $A = (ab) \in \6$, $a, b \in \4$, $a< b$:
\beq
{\CalN}_{A} = \sum_{\bnu \in {\Gamav}} \sum_{{\alpha} \in [n_{A, \bnu}]} e^{{\beta}{\ac}_{A, \bnu, \alpha}} {\CalL}_{\bnu} = \sum_{{\bnu} \in {\Gamav}} N_{A, \bnu}({\beta})  {\CalL}_{\bnu}
\eeq

\item{} The string ${\bqe} = ({\qe}_{\bnu})_{\bnu \in {\Gammadi}}$ of $|{\Gamab}| = \prod_{\kappa} p_{\kappa}^{l_{\kappa}}$ fugacities
\beq
{\qe}_{\bnu} \in {\BC}\, , \quad {\bnu} \in {\Gamav}
\eeq
obeying $|{\qe}_{\bnu} | < 1$.

\end{enumerate}
Define, for $S \subset {\4}$:
\beq
P_{S, \bnu}({\beta}) = \sum_{J \subset S} \, \prod_{a \in J} \left( - e^{{\beta}{\ve}_{a}} \right)
 \, {\delta}_{\Gamma^{\vee}} \left( - \bnu + \sum_{a \in J} {\bro}_{a} \right) \ ,  \qquad
 {\CalP}_{S} = \sum_{\bnu \in {\Gamma}^{\vee}} P_{S, {\bnu}} \, {\CalL}_{{\bro}_{S}} \label{eq:abps} \eeq

Define, for $\bla \in {\Lambda}^{{\bN}_{\Gamab}}$, $A \in \6$, $A = (ab)$ as before:
\begin{multline}
{\CalK}_{A} =  \sum_{\bnu \in {\Gamab}^{\vee}} \sum_{{\alpha} \in [n_{A, \varpi}]} \sum_{(i,j) \in {\lambda}^{(A, {\bnu}, {\alpha})}} e^{{\beta}\left({\ac}_{A, \bnu, \alpha} + {\ve}_{a}(i-1)+ {\ve}_{b}(j-1)\right)} \, {\CalL}_{\bnu + {\bro}_{a}(i-1) + {\bro}_{b}(j-1)}\, = \\
\sum_{\bnu \in {\Gamav}} K_{A, \bnu} ({\beta}) {\CalL}_{\bnu}
\end{multline}

\subsubsec{The abelian orbifold model statistical weights}

Define,  for  $A = \{ a, b \}$, $a< b \in {\4}$, cf. \eqref{eq:abps} 
\beq
{\CalT}_{A} \ = \ \sum_{{\bnu} \in {\Gamav}} T_{A, \bnu}  {\CalL}_{\bnu} \,
 = \, {\CalN}_{A} {\CalK}_{A}^{*} + q_{A} {\CalK}_{A} {\CalN}_{A}^{*} \otimes {\CalL}_{{\bro}_{A}}  - {\CalK}_{A} {\CalK}_{A}^{*} {\CalP}_{A} \ . 
\eeq
The statistical weight of $\bla$ is given by the following expression:
\beq
Z_{\bla}^{\Gamab} =\  \left( \prod_{{\bnu} \in {\Gamav}} \, {\qe}_{\bnu}^{k_{\bnu}}\right) \  {\ep} \left[ - T_{\bla}^{\Gamab} \right] 
\label{eq:swbarlam2}
\eeq
where
\beq
k_{\bnu} = \sum_{A \in \6, {\alpha} \in [n_{A, {\bnu}}]} | {\lambda}^{(A, {\bnu}, {\alpha})} | \quad , 
\eeq
\beq
T_{\bla}^{\Gamab} = \sum_{A \in {\6},\, {\bnu} \in {\Gamav}} \left(
P_{{\varphi}(A), -\bnu} T_{A, \bnu} + \sum_{{\bnu}' \in {\Gamav}} \left( P_{{\bar A}, {\bnu}} N_{A, \bnu'} \left( \sum_{B\neq A} K_{B, \bnu+\bnu'} \right)^{*}  \, - \, 
P_{\4, \bnu} K_{A, \bnu'}  \left( \sum_{B>A} K_{B, \bnu+\bnu'} \right)^{*}  \right)\right)  \, , 
\label{eq:tlamgam}
\eeq
and the full abelian orbifold partition function is defined as
\beq
{\CalZ}^{\Gamab}_{\rm spiked} ({\tilde\ac}, {\bro}, {\bar\ep}; {\tilde\qe})  = Z^{\rm pert, \Gamab} ({\tilde\ac}, {\bro}, {\bar\ve}) \, \sum_{\bla} Z_{\bla}^{\Gamab}
\label{eq:abelorb}
\eeq
where the prefactor is given by the following formulas:

\subsubsec{The abelian orbifold gauge origami perturbative factors}

Define:

\begin{multline}
Z^{\rm pert, \Gamab} ({\tilde\ac}, {\bro}, {\bar\ve})  =  \prod\limits_{A \in {\6}} \ Z_{{\CalN}=2*}^{{\rm pert}, A, \Gamab} ( {\bf\ac}_{A},  {\bro}_{a}, {\bro}_{b}, {\bar\ve} ) \ \times \\ 
 \prod\limits_{\{a , b, c \} \subset {\4}}\, Z^{{\rm pert}, a|bc, \Gamab}_{\rm fold} 
 ({\bf\ac}_{ab}, {\bf\ac}_{ac},  {\bro}_{a}, {\bro}_{b}, {\bro}_{c}, {\bar\ve} ) \  \times \\
  \prod\limits_{A \in {\6}, A< {\bar A}} \, Z^{{\rm pert}, A, \Gamab}_{\rm cross}
( {\bf\ac}_{A}, {\bf\ac}_{\bar A} ,  {\bro}, {\bar\ve}) \end{multline}
where ($A = ab$):
\beq
Z_{{\CalN}=2*}^{{\rm pert}, A, \Gamab} ( {\bf\ac}_{A},  {\bro}, {\bar\ve} ) = 
\prod_{{\bnu}, {\bnu}' \in {\Gamav}}
\prod_{{\alpha} \in [n_{A, {\bnu}}] , {\alpha'} \in [n_{A, {\bnu}'}]} \frac{{\Gamma}_{2,\Gamab} \left(\begin{matrix} & {\ac}_{A, {\bnu}, \alpha} - {\ac}_{A, {\bnu}' , \alpha'}  & ; & {\ve}_{a}, {\ve}_{b} \\
& {\bnu} - {\bnu}'  & ;  & {\bro}_a, {\bro}_b  \\ \end{matrix} \ \right)}{{\Gamma}_{2,\Gamab} \left(\begin{matrix} & {\ac}_{A, {\bnu}, \alpha} - {\ac}_{A, {\bnu}' , \alpha'} + {\ve}_{\varphi (A)} & ; & {\ve}_{a}, {\ve}_{b} \\
& {\bnu} - {\bnu}' + {\bro}_{\varphi (A)} & ;  & {\bro}_a, {\bro}_b  \\ \end{matrix} \ \right)} \label{eq:orbin2*}
\eeq
where the projected double gamma
\beq
{\Gamma}_{2, \Gamab} \left(\begin{matrix} & x & ; & y', y'' \\
& {\bnu} & ;  & {\bro}', {\bro}''  \\ \end{matrix} \ \right)
 \sim
\prod_{i, j \geq 1} \left( x + y' (i-1) + y''(j-1) \right)^{{\delta}_{\Gamav} \left( \bnu + {\bro}'(i-1) + {\bro}''(j-1) \right)}
\eeq
can be easily expressed in terms of the ordinary $\Gamma_2$'s, 

$\bullet$

\beq
Z^{{\rm pert}, a|bc,  \Gamab}_{\rm fold} ({\bf\ac}_{ab}, {\bf\ac}_{ac}, \rho, {\bar\ve} ) =  \ \prod_{{\bnu}, {\bnu}' \in {\Gamav}}\, \prod_{{\alpha} \in [n_{ab, \bnu}]}
\prod_{{\beta} \in [n_{ac, \bnu'}]}\
{\Gamma}_{1,\Gamab} \left(\begin{matrix} & {\ac}_{ab, \bnu, \alpha} - {\ac}_{ac, \bnu', \beta} + {\ve}_{a} + {\ve}_{c}  & ; & {\ve}_{a} \\
& {\bnu} - {\bnu}' + {\bro}_{a} + {\bro}_{c} & ;  & {\bro}_a  \\ \end{matrix} \ \right) 
\eeq
where
\beq
{\Gamma}_{1, \Gamab} \left(\begin{matrix} & x  & ; & y \\
& {\bnu}  & ;  & {\bro}  \\ \end{matrix} \ \right)
 \sim \prod_{i=1}^{\infty}
\left( x  + y(i-1) \right)^{{\delta}_{\Gamav}({\bnu} + {\bro}(i-1))} 
\eeq
can be easily expressed in terms of the ordinary gamma-functions, 

$\bullet$

\beq
Z^{{\rm pert}, A, \Gamab}_{\rm cross}
( {\bf\ac}_{A}, {\bf\ac}_{\bar A} ,\rho,  {\bar\ve}) \, = \, \prod_{{\bnu}, {\bnu}' \in {\Gamav}}\, \prod_{{\alpha} \in [n_{A,\bnu}]} \prod_{{\beta} \in [n_{B, \bnu'}]} \ \left( {\ac}_{A, \bnu, \alpha} - {\ac}_{{\bar A}, \bnu', \beta} +
 {\ve}_{\bar A} \right)^{{\delta}_{\Gamav}({\bnu}-{\bnu}' - {\bro}_{A})}  
 \label{eq:pertcross}
 \eeq

\subsec{The {\bf abelian} $\times$ {\bf ALE} case}

Fix $\Gamab$, $\gamma$ of $D$ or $E$ type. Let ${\Gammadi} = {\Gamab} \times {\Gamma}_{\gamma}$. The irreps of $\Gammadi$ are labelled by the pairs ${\bom} = ({\bnu}, {\ib})$, ${\bnu} \in {\Gamav}, {\ib} \in \Vg$. 
Fix the discrete data: the dimension vectors ${\bn} = (n_{\bom})_{{\bom} \in {\Gammadi}^{\vee}}$,   ${\bw} = ( w_{\bom})_{{\bom} \in {\Gammadi}^{\vee}}$, and the two characters ${\bro}_{l}, {\bro}_{r} : {\Gamab} \to U(1)$.
The orbifold gauge origami in this case depends on the following continuous data:

\begin{enumerate}

\item
Two complex numbers ${\ve}_{1}, {\ve}_{2}$, and ${\ve} = {\ve}_{1} + {\ve}_{2}$. 
\item
Two sets of Coulomb parameters: ${\ba} = ( {\ac}_{{\bnu}, {\ib},\alpha}) \in {\BC}^{{\bN}^{+}_{\Gammadi}}$, ${\bb} = ( {\fb}_{{\bnu}, {\ib}, {\beta}}) \in {\BC}^{{\bN}^{-}_{\Gammadi}}$
\beq
{\ac}_{{\bnu}, {\ib},\alpha} \in {\BC}, \qquad {\alpha} \in [n_{{\bnu}, {\ib}}], \qquad {\fb}_{{\bnu}, {\ib}, {\beta}} \in {\BC}, \qquad {\beta} \in [w_{{\bnu}, {\ib}}]
\label{eq:acfb}
\eeq
\item
The string ${\bf\qe} = ({\qe}_{\bom})_{\bom \in {\Gammadi}^{\vee}}$ of $\vert {\Gammadi} \vert$ fugacities:
\beq
{\qe}_{\bom} \in {\BC}\, , \qquad | {\qe}_{\bom} | < 1 
\eeq

\end{enumerate}
The geometric action ${\Gammadi} \to SU(4)$ defines the following three representations:
\beq
{\BC}^{1}_{1} \equiv {\CalL}_{{\bro}_{l}+{\bro}_{r}}, \quad {\BC}^{1}_{2} \equiv {\CalL}_{{\bro}_{l} - {\bro}_{r}}, \qquad
{\BC}^{2}_{34} = {\CalL}_{-{\bro}_l} \otimes {\bf 2}  \ , 
\eeq
which obey
\beq
{\BC}^{1}_{1} \otimes {\BC}^{1}_{2} \otimes {\Lambda}^{2} {\BC}^{2}_{34} = {\CalR}_{\bf 0}\, , 
\label{eq:trivdet}
\eeq
the trivial representation. 
Write ${\CalP}_{\4} = {\CalP}_{12} {\CalP}_{34}$, with:
\beq
{\CalP}_{12} = 1 - q_{1} {\BC}^{1}_{1} - q_{2} {\BC}^{1}_{2} + 
e^{{\beta}{\ve}} {\CalL}_{2 \bro_{l}} \ , \ 
{\CalP}_{34} =  1 - q^{-\frac 12} {\BC}^{2}_{34}  + q^{-1} {\CalL}_{-2{\bro}_{l}}\ . 
\label{eq:p12p34abale}
\eeq
Finally, define 
\beq
{\CalN} = \sum_{\bom \in {\Gammadi}^{\vee}}  \sum_{{\alpha} \in [n_{\bom}]} e^{{\beta}{\ac}_{\bom, \alpha}} \  {\CalR}_{\bom}
\, , \qquad
{\CalW} = \sum_{\bom \in {\Gammadi}^{\vee}} \sum_{{\tilde\alpha} \in [w_{\bom}]} e^{{\beta}{\fb}_{\bom, {\tilde\alpha}}} \  {\CalR}_{\bom}
\eeq
$\bullet$
The random variables ${\bla}_{{\rm ab} \times {\rm ale}}$ in the $\Gammadi$-orbifold gauge origami model were defined in 
\eqref{eq:abalec}. The statistical weight of ${\bla}_{{\rm ab} \times {\rm ale}}$ is an integral over the product of Nakajima-Young varieties:
\beq
{\CalX}_{{\bla}_{{\rm ab} \times {\rm ale}}} = 
\varprod_{{\bnu} \in {\Gamav}} \varprod_{{\ib} \in {\Vg}}  \varprod_{{\tilde\alpha} \in [w_{{\bnu}, {\ib}}]} \ {\bY}^{\ib}_{\gamma} \left( {\mu}_{{\bnu}, \ib,  {\tilde\alpha}} \right) \ ,
\qquad {\mu}_{{\bnu}, \ib, {\tilde\alpha}} \in {\Lambda}^{\ib}_{\gamma}
\label{eq:compny}
\eeq
Define
\beq
{\CalK} = \sum_{{\bnu} \in  {\Gamav}, {\ib} \in {\Vg}, {\alpha} \in [n_{\bnu, \ib}]}  \sum_{(i,j) \in {\lambda}^{({\ib}, {\varpi}, {\alpha})}} e^{{\beta}{\ac}_{\bnu, {\ib}, {\alpha}}} q_{1}^{i-1}
q_{2}^{j-1} \, {\CalL}_{{\bnu} + {\bro}_{l}(i+j-2) + {\bro}_{r}(i-j)} \otimes R_{\ib}\ ,  
\eeq
\beq
{\CalT}_{12} = {\CalN}{\CalK}^{*} + {\CalN}^{*}{\CalK}q {\CalL}_{2\bro_{l}} - {\CalP}_{12} {\CalK}{\CalK}^{*} = \sum_{{\bom} \in {\Gammadi}^{\vee}} {\CalT}_{12,\bom} \, {\CalR}_{\bom}
\eeq
so that, in particular
\begin{multline}
{\CalT}_{12,\bnu, {\bf 0}} = \sum_{{\ib} \in \Vg} \sum_{{\bnu}' \in {\Gamav}} \ \left( 
{\CalN}_{{\bnu}+{\bnu}', {\ib}} {\CalK}_{{\bnu}', {\ib}}^{*} + q {\CalN}_{2{\bro}_{l}+{\bnu}', {\ib}}^{*} {\CalK}_{{\bnu}+{\bnu}', {\ib}} - {\CalK}_{{\bnu}+{\bnu}', {\ib}}{\CalK}_{{\bnu}', {\ib}}^{*}  \right. \\
\left. + q_{1} {\CalK}_{{\bnu}+{\bnu}', {\ib}}{\CalK}_{{\bro}_{l} + {\bro}_{r}+ {\bnu}', {\ib}}^{*}
+ q_{2} {\CalK}_{{\bnu}+{\bnu}', {\ib}}{\CalK}_{{\bro}_{l} - {\bro}_{r}+ {\bnu}', {\ib}}^{*} 
+ q  {\CalK}_{{\bnu}+{\bnu}', {\ib}}{\CalK}_{2{\bro}_{l}+{\bnu}', {\ib}}^{*} \right) \ .
\end{multline}
Define:
\beq
{\CalV} = \sum_{{\ib}, {\tilde\ib} \in {\Vg}} \sum_{{\bnu} \in  {\Gamav}} \sum_{{\tilde\alpha} \in [w_{\bnu, \ib}]} \sum_{n \geq 0}
e^{{\beta}{\fb}_{\bnu, \ib, {\tilde\alpha}}} q^{- \frac n2}\, {\rm Ch}\left( V_{{\tilde\ib} , n}^{\ib}( {\mu}_{{\bnu}', {\ib}', {\tilde\alpha}})  \right) \, {\CalL}_{\bnu -  n {\bro}_{l}}  \otimes R_{\tilde\ib} 
\eeq
\beq
{\CalT}_{34} = {\CalW}{\CalV}^{*} + q^{-1} \,{\CalW}^{*}{\CalV} {\CalL}_{-2\bro_{l}} - {\CalP}_{34} {\CalV}{\CalV}^{*} = \sum_{{\bom} \in {\Gammadi}^{\vee}} {\CalT}_{34,\bom} \, {\CalR}_{\bom}
\eeq
{}We view $\CalN, \CalW, \CalK, \CalV, \CalT_{12}, \CalT_{34}$ as the
 $K[ T_{{\Hf}_{\Gammadi}} ] \otimes K({\Gammadi})$-valued linear combinations of Chern characters of vector bundles over ${\CalX}_{{\bla}_{{\rm ab} \times {\rm ale}}}$, as well as ${\CalP}_{1} = 1 - q_{1} {\BC}^{1}_{1}, \ {\CalP}_{2} = 1 - q_{2} {\BC}^{1}_{2}$.  

{}The measure \eqref{eq:swbarlam} dressed with a partial perturbative contribution, the orbifold version of \eqref{eq:pertcross}, is generalized to 
\beq
{\CalZ}^{\rm pert}_{\Gammadi, \rm cross} z_{{\bla}_{{\rm ab}\times{\rm ale}}} = \left( \prod_{\bom \in {\Gammadi}^{\vee}}  {\qe}^{k_{\bom}} \right) \  \int_{{\CalX}_{{\bla}_{{\rm ab} \times {\rm ale}}}} {\ep} \left[ - \left[{\CalR}_{0} \right]{\CalT}_{{\bla}_{{\rm ab}\times{\rm ale}}} + \left[L_{\bf 0} {\CalR}_{0} \right]{\CalT}_{{\bla}_{{\rm ab}\times{\rm ale}}} \right] \, , 
\label{eq:swbarlamorb}
\eeq
where (cf. \eqref{eq:tny}):
\begin{multline}
 \left[{\CalR}_{0} \right]{\CalT}_{{\bla}_{{\rm ab}\times{\rm ale}}} =  \left[{\CalR}_{0} \right] \left(   - q\,  {\CalL}_{2\bro_{l}}\, {\CalN}^{*} {\CalW}+
  {\CalT}_{12} + {\CalP}_{34} {\CalN} {\CalV}^{*} + 
 {\CalP}_{1} {\CalT}_{34} + {\CalP}_{12} {\CalW} {\CalK}^{*}  -  {\CalP}_{\4} {\CalK}{\CalV}^{*} \right) \, -  \\
 \\
\qquad  - q^{-\frac 12} \left[{\CalR}_{0} \right] \sum_{\bnu \in {\Gamav}} \sum_{e \in \Eg} \left( {\CalN}_{{\bnu} + {\bro}_{l}, t(e)} 
 {\CalK}_{{\bnu}, s(e)}^{*} - {\CalK}_{{\bnu}+{\bro}_{l}, t(e)} {\CalK}_{{\bnu}, s(e)}^{*}
 - q\, {\CalK}_{{\bnu}-{\bro}_{l}, t(e)} {\CalK}_{{\bnu}, s(e)}^{*} + \right. \\
 \qquad\qquad \qquad\qquad\qquad\qquad \left. + 
 q_{1} {\CalK}_{{\bnu}-{\bro}_{r}, t(e)} {\CalK}_{{\bnu}, s(e)}^{*}
 + q_{2} {\CalK}_{{\bnu}+{\bro}_{r}, t(e)} {\CalK}_{{\bnu}, s(e)}^{*}\right)  \, , \\
  \\
 [L_{\bf 0} {\CalR}_{\bf 0}] {\CalT}_{{\bla}_{{\rm ab}\times{\rm ale}}} = 
 [L_{\bf 0}{\CalR}_{\bf 0}] {\CalT}_{34}   = 
 {\CalT}_{{\CalX}_{{\bla}_{{\rm ab}\times{\rm ale}}}} \\
\label{eq:tlamgamale}
\end{multline}
and $[{\CalR}_{0}]({\ldots})$ denotes taking the $\Gammadi$-invariant part in $({\ldots})$, i.e. the contribution of the trivial representation of $\Gammadi$, while $[L_{\bf 0}{\CalR}_{0}]{\CalT}$ (cf. \eqref{eq:irrept})
denotes the $T_{{\Hf}_{\Gammadi}} \times {\Gammadi}$-invariant part. Geometrically, the $T_{{\Hf}_{\Gammadi}} \times {\Gammadi}$-invariant $[L_{\bf 0}{\CalR}_{\bf 0}] {\CalT}_{34} $ is the tangent space to the variety ${\CalX}_{{\bla}_{{\rm ab}\times{\rm ale}}}$ so its contribution is subtracted from the measure as the rest is being integrated over ${\CalX}_{{\bla}_{{\rm ab}\times{\rm ale}}}$ (note that $L_{\bf 0}{\CalR}_{\bf 0} = L_{\bf 0}R_{\bf 0}$ as ${\Gamab}$ action is contained in $T_{{\Hf}_{\Gammadi}}$). Finally, 
\begin{multline}
k_{\bnu, {\ib}} =  \sum_{{\bnu}'\in {\Gamav}} \left( 
\sum_{{\alpha} \in [n_{\bnu', \ib}]} \sum_{(i,j) \in {\lambda}^{({\bnu}', {\ib}, {\alpha})}}\
{\delta}_{{\Gamav}} \left( {\bnu}' + {\bro}_{l}(i+j-2) + {\bro}_{r}(i-j) - {\bnu} \right)  + \right. \\
+ \left. 
\sum_{{\ib}' \in {\Vg}} \sum_{{\tilde\alpha} \in [w_{\bnu', \ib'}]} \sum_{n\geq 0} \
{\delta}_{{\Gamav}} \left( {\bnu}' - n {\bro}_{l} - {\bnu} \right) {\nu}^{{\ib}'}_{{\ib}, n} ( {\mu}_{{\bnu}', {\ib}', {\tilde\alpha}}) \right)
\label{eq:instom}
\end{multline}

Of course, this formalism also applies to $\gamma = {\hat A}_{k}$. In this case the formulas \eqref{eq:instom}, \eqref{eq:swbarlamorb} reduce to the ${\ve}_{3} = {\ve}_{4}$ limit
of the {\bf abelian} orbifold case of crossed instantons \cite{Nekrasov:2016qym}. 
 
\subsec{The {\bf ALE} $\times$ {\bf ALE} case}

Fix $\Gamab$, and two quivers $\gamma' , \gamma''$of $D$ or $E$ type. 
In this section $\Gammadi = \Gamab \times {\Gamma}_{\gamma'} \times {\Gamma}_{\gamma''}$, with its irreps $\bom = ( {\bnu}, {\ib}', {\ib}'' )$, ${\bnu} \in \Gamav$, ${\ib}' \in {\rm Vert}_{\gamma'}$, ${\ib}''  \in {\rm Vert}_{\gamma''}$. 

Fix the discrete data: the dimension vectors ${\bn} = (n_{\bom})_{{\bom} \in {\Gammadi}^{\vee}}$,   ${\bw} = ( w_{\bom})_{{\bom} \in {\Gammadi}^{\vee}}$, one character ${\bro}: {\Gamab} \to U(1)$, equivalently a representation ${\CalL}_{\bro} \in {\Gamav}$. 
The orbifold gauge origami in this case depends on the following continuous data:

\begin{enumerate}

\item
A complex number  ${\ve} \in {\BC}$. 
\item
Two sets of Coulomb parameters (cf. \eqref{eq:nlababaleale}): 
\beq
{\ba} = ( {\ac}_{{\bom}, {\alpha}} ) \in {\BC}^{{\bN}_{\Gammadi}^{+}}, \qquad {\bb} = ({\fb}_{{\bom}, {\beta}}) \in {\BC}^{{\bN}_{\Gammadi}^{-}} \ ,
\label{eq:acfb}
\eeq 
where ${\ac}_{\bom, \alpha} \in {\BC}$, ${\alpha} \in [n_{\bom}]$, ${\fb}_{\bom, {\beta}} \in {\BC}$, ${\beta} \in [w_{\bom}]$.
\item
The string ${\bf\qe} = ({\qe}_{\bom})_{\bom \in {\Gammadi}^{\vee}}$ of $\vert {\Gammadi} \vert$ fugacities:
\beq
{\qe}_{\bom} \in {\BC}\, , \qquad | {\qe}_{\bom} | < 1 
\eeq

\end{enumerate}
The geometric action ${\Gammadi} \to SU(4)$ defines the following two representations:
\beq
{\BC}^{2}_{12} \equiv {\CalL}_{\bro} \otimes {\bf 2}', \qquad
{\BC}^{2}_{34} = {\CalL}_{-\bro} \otimes {\bf 2}'' 
\eeq
Define:
\beq
{\CalP}_{12} = 1 - q^{\frac 12} {\BC}^{2}_{12} + q {\CalL}_{2\bro} \ , \qquad
{\CalP}_{34} = 1 - q^{-\frac 12} {\BC}^{2}_{34} + q^{-1} {\CalL}_{-2\bro}
 \eeq
 $\bullet$
 The random variables ${\bla}_{{\rm ale} \times {\rm ale}}$ in the $\Gammadi$-orbifold gauge origami model were defined 
 in \eqref{eq:alealec}. The statistical weight of ${\bla}_{{\rm ale} \times {\rm ale}}$
 is given by the integral over the product
 of Nakajima-Young varieties
 \beq
 {\CalX}_{{\bla}_{{\rm ale} \times {\rm ale}}} = 
 \varprod_{\bnu \in {\Gamav}} \varprod_{{\ib}'\in {\rm Vert}_{\gamma'}, {\ib}'' \in {\rm Vert}_{\gamma''}} \, \left( \varprod_{{\alpha} \in [n_{{\bnu}, {\ib}', {\ib}''}]} {\bY}^{\ib'}_{\gamma'} ( {\mu}^{({\bnu}, {\ib}', {\ib}''; {\alpha})} ) \, \times \, \varprod_{{\beta} \in [w_{{\bnu}, {\ib}', {\ib}''}]} {\bY}^{\ib''}_{\gamma''} ( {\tilde\mu}^{({\bnu}, {\ib}', {\ib}''; {\beta})} ) \right)
 \label{eq:alealecc}
 \eeq
with ${\mu}^{({\bom}, {\alpha})} \in {\Lambda}^{\ib'}_{\gamma'}$, ${\tilde\mu}^{({\bom}, {\beta})} \in {\Lambda}^{\ib''}_{\gamma''}$.
Define
\beq
{\CalN} = \sum\limits_{{\bnu} \in {\Gamav}, {\ib}' \in {\rm Vert}_{\gamma'},  
{\ib''} \in {\rm Vert}_{\gamma''}} \sum_{{\alpha} \in [n_{{\bnu}, {\ib}', {\ib}''}]} 
  e^{{\beta}{\ac}_{{\bnu}, {\ib}', {\ib}'';  {\alpha}}}  \, {\CalL}_{\bnu} 
\otimes R_{\ib'}' \otimes R_{\ib''}''  \ ,  \eeq
\begin{multline}
{\CalK} = \sum\limits_{\tiny \begin{matrix}
{\ib}, {\ib}' \in {\rm Vert}_{\gamma'} \\ 
{\ib''} \in {\rm Vert}_{\gamma''} \\  
{\bnu} \in  {\Gamab}^{\vee} \end{matrix}} \
\sum\limits_{\tiny\begin{matrix} n \geq 0 \\
{\alpha} \in [n_{{\bnu}, \ib,  {\ib''}}] \end{matrix}}  
\  e^{{\beta}{\ac}_{{\bnu}, {\ib}, {\ib}'';  {\alpha}}}  q^{\frac n2} 
{\rm Ch} \left( V_{{\ib}, n}^{\ib'} \left( {\mu}^{({\bnu}, {\ib}', {\ib}''; {\alpha})}  \right) \right) {\CalL}_{{\bnu} +  n \bro} 
\otimes R_{\ib'}' \otimes R_{\ib''}''   = \\
\sum_{\bom \in {\Gammadi}^{\vee}} {\CalK}_{\bom} \, {\CalR}_{\bom}\ ,  \end{multline}
\beq
{\CalT}_{12} = {\CalN}{\CalK}^{*} + {\CalN}^{*}{\CalK} q{\CalL}_{2\bro} - {\CalP}_{12}{\CalK}{\CalK}^{*} \, , \eeq
and
\beq
{\CalW} = \sum\limits_{{\bnu} \in {\Gamav}, {\ib}' \in {\rm Vert}_{\gamma'} , {\tilde\ib}'' \in {\rm Vert}_{\gamma''}} 
\sum\limits_{{\tilde\alpha} \in [w_{{\bnu}, \ib',  {\ib}''}]}  
e^{{\beta}{\fb}_{{\bnu}, \ib', {\ib''};  {\tilde\alpha}}}  \, {\CalL}_{\bnu} 
 \otimes R_{\ib'}' \otimes R_{\ib''}'' \ ,  \eeq
\begin{multline}
{\CalV} = \sum\limits_{\tiny \begin{matrix}
{\ib}' \in {\rm Vert}_{\gamma'} \\ 
{\tilde\ib}, {\ib}'' \in {\rm Vert}_{\gamma''} \\  
{\bnu} \in  {\Gamav} \end{matrix}} \
\sum\limits_{\tiny\begin{matrix} n \geq 0 \\
{\tilde\alpha} \in [w_{{\bnu}, \ib,  {\ib''}}] \end{matrix}}  
e^{{\beta}{\fb}_{{\bnu}, \ib, {\tilde\ib};  {\tilde\alpha}}} \, q^{-\frac n2}\, {\rm Ch} 
\left( {\tilde V}_{{\tilde\ib} , n}^{\ib''} \left( {\tilde\mu}^{({\bnu}, {\ib}', {\ib}''; {\beta})}  \right) \right) \, {\CalL}_{\bnu -  
n \bro} 
 \otimes R_{\ib'}' \otimes R_{\tilde\ib}''  = \\
\sum_{\bom \in {\Gammadi}^{\vee}} {\CalV}_{\bom} \, {\CalR}_{\bom}\ ,  \end{multline}
 \beq
{\CalT}_{34} = {\CalW}{\CalV}^{*} + {\CalW}^{*}{\CalV} q^{-1}{\CalL}_{-2\bro} - {\CalP}_{34}{\CalV}{\CalV}^{*} \,  ,\\
\eeq
a $K[ T_{{\Hf}_{\Gammadi}} ] \otimes K({\Gammadi})$-valued linear combination of vector bundles over ${\CalX}_{{\bla}_{{\rm ale} \times {\rm ale}}}$, 
where the vector bundles over ${\CalX}_{{\bla}_{{\rm ale} \times {\rm ale}}}$ denoted with some abuse of notation by $V_{{\ib}', n}^{\ib}\left( {\mu}^{({\bnu}, {\ib}', {\ib}''; {\alpha})}  \right)$, ${\tilde V}_{{\tilde\ib}'' , n}^{\tilde\ib}\left( {\tilde\mu}^{({\bnu}, {\ib}', {\ib}''; {\beta})}  \right)$  are the pullbacks of the bundles  $V_{{\ib}', n}^{\ib} \left( {\mu}^{({\bnu}, {\ib}', {\ib}''; {\alpha})}  \right)  \to {\bY}^{\ib}_{\gamma'} \left( {\mu}^{({\bnu}, {\ib}', {\ib}''; {\alpha})}  \right)$, $V_{{\tilde\ib}'' , n}^{\tilde\ib} 
\left( {\tilde\mu}^{({\bnu}, {\ib}', {\ib}''; {\beta})}  \right) \to {\bY}^{\ib}_{\gamma''} \left( {\tilde\mu}^{({\bnu}, {\ib}', {\ib}''; {\beta})}  \right)$ under the projections to the respective factors in \eqref{eq:alealecc}. 

The measure \eqref{eq:swbarlam} dressed with a partial perturbative contribution, the $\Gammadi$-orbifold version of \eqref{eq:pertcross}, is now generalized to 
\beq
{\CalZ}^{\rm pert}_{\Gammadi, \rm cross} z_{{\bla}_{{\rm ale}\times{\rm ale}}} = \left( \prod_{\bom \in {\Gammadi}^{\vee}}  {\qe}^{k_{\bom}} \right) \  \int_{{\CalX}_{{\bla}_{{\rm ale}\times{\rm ale}}}} \ {\ep} \left[ - \left[{\CalR}_{0} \right]{\CalT}_{{\bla}_{{\rm ale}\times{\rm ale}}} + \left[L_{\bf 0} {\CalR}_{0} \right]{\CalT}_{{\bla}_{{\rm ale}\times{\rm ale}}} \right] \, , 
\label{eq:zalealeorb}
\eeq
where (cf. \eqref{eq:tny}):
\begin{multline}
 \left[{\CalR}_{0} \right]{\CalT}_{{\bla}_{{\rm ale}\times{\rm ale}}} =  \left[{\CalR}_{0} \right] \left(   - q\,  {\CalL}_{2\bro}\, {\CalN}^{*} {\CalW}+
  {\CalT}_{12} + {\CalP}_{34} {\CalN} {\CalV}^{*} + 
  {\CalT}_{34} + {\CalP}_{12} {\CalW} {\CalK}^{*}  -  {\CalP}_{\4} {\CalK}{\CalV}^{*} \right) \, - 
 \\  -  \left[{\CalR}_{0} \right] \left( q^{-\frac 12} {\BC}^{2}_{34} {\CalN}{\CalK}^{*} + q^{\frac 12} {\BC}^{2}_{12} {\CalW}{\CalV}^{*} - q^{-\frac 12} {\BC}^{2}_{34} {\CalK}{\CalK}^{*} - 
 q^{\frac 12} {\BC}^{2}_{12} {\CalV}{\CalV}^{*} \right) \, -  \\
 - \sum_{\bnu \in {\Gamav}} \sum_{e' \in {\rm Edge}_{\gamma'}}\sum_{e'' \in {\rm Edge}_{\gamma''}} \left(  {\CalK}_{{\bnu}, t(e'), s(e'')} {\CalK}_{{\bnu}, s(e'), t(e'')}^{*} + {\CalK}_{{\bnu}, t(e'), t(e'')} {\CalK}_{{\bnu}, s(e'), s(e'')}^{*} \right) 
 \\
 - \sum_{\bnu \in {\Gamav}} \sum_{e' \in {\rm Edge}_{\gamma'}}\sum_{e'' \in {\rm Edge}_{\gamma''}} \left(  {\CalV}_{{\bnu}, s(e'), t(e'')} {\CalV}_{{\bnu}, t(e'), s(e'')}^{*} + {\CalV}_{{\bnu}, t(e'), t(e'')} {\CalV}_{{\bnu}, s(e'), s(e'')}^{*} \right) \
 , \\
 [L_{\bf 0} {\CalR}_{\bf 0}] {\CalT}_{{\bla}_{{\rm ale}\times{\rm ale}}} = 
 [L_{\bf 0}{\CalR}_{\bf 0}] \left( {\CalT}_{12}+{\CalT}_{34} \right)   = 
 {\CalT}_{{\CalX}_{{\bla}_{{\rm ale}\times{\rm ale}}}} \\
\label{eq:tlamgamale}
\end{multline}
and
\begin{multline}
k_{{\bnu}, {\ib}, {\tilde\ib}} =  \sum_{{\bnu'}\in {\Gamav}} \sum_{n\geq 0} \left( 
\sum_{{\ib}' \in {\rm Vert}_{\gamma'}} \sum_{{\alpha} \in [n_{{\bnu'}, \ib', {\tilde\ib}}]} 
\
{\delta}_{{\Gamav}} \left( {\bnu'} + n {\bro} - {\bnu} \right)  \,
{\nu}^{\ib'}_{{\ib}, n} ({\mu}_{{\bnu'}, {\ib}', {\tilde\ib}; {\alpha}}) + \right. 
\\
+ \left. 
\sum_{{\tilde\ib}'' \in {\rm Vert}_{\gamma''}} \sum_{{\tilde\alpha} \in [w_{{\bnu'}, \ib, {\tilde\ib}''}]}  \
{\delta}_{{\Gamav}} \left( {\bnu'} -  n {\bro} - {\bnu} \right) \, 
{\nu}^{{\tilde\ib}''}_{{\tilde\ib}, n} ( {\tilde\mu}_{{\bnu}',{\ib}, {\tilde\ib}''; {\tilde\alpha}}) \right)
\label{eq:instomaleale}
\end{multline}
To compute the measure \eqref{eq:instomaleale} we use \eqref{eq:mckay} to write:
\begin{multline}
\left[{\CalR}_{0} \right] \left( {\bf 2}' \otimes {\bf 2}'' \otimes  {\CalK}{\CalK}^{*}  \right) = 
\sum_{\bnu \in {\Gamav}} \sum_{e' \in {\rm Edge}_{\gamma'}}\sum_{e'' \in {\rm Edge}_{\gamma''}} \left( {\CalK}_{{\bnu}, s(e'), s(e'')} {\CalK}_{{\bnu}, t(e'), t(e'')}^{*} + 
{\CalK}_{{\bnu}, t(e'), s(e'')} {\CalK}_{{\bnu}, s(e'), t(e'')}^{*} + \right. \\
\left. {\CalK}_{{\bnu}, s(e'), t(e'')} {\CalK}_{{\bnu}, t(e'), s(e'')}^{*} + {\CalK}_{{\bnu}, t(e'), t(e'')} {\CalK}_{{\bnu}, s(e'), s(e'')}^{*} \right) 
\end{multline}
and similarly for ${\bf 2}' \otimes {\bf 2}'' \otimes {\CalV} {\CalV}^{*}$. To compute, e.g. 
the contribution $[ {\CalR}_{\bf 0} ] \left( {\CalP}_{12} {\CalW}{\CalK}^{*} \right)$ to \eqref{eq:instomaleale} we also use \eqref{eq:mckay}:
\begin{multline}
\left[ {\CalR}_{\bf 0} \right] \left( {\CalP}_{12} {\CalW}{\CalK}^{*} \right) = \\
\sum_{\bnu \in {\Gamav}}  \sum_{{\ib}'' \in {\rm Vert}_{\gamma''}} \left\{ 
\sum_{{\ib}' \in {\rm Vert}_{\gamma'}}  \left( {\CalW}_{{\bnu}, {\ib}', {\ib}''}  {\CalK}_{{\bnu}+{\bro}, {\ib}', {\ib}''}^{*} +
q {\CalW}_{{\bnu}, {\ib}', {\ib}''} {\CalK}_{{\bnu}+2{\bro}, {\ib}', {\ib}''}^{*} \right) - \right. \\
\left. \qquad\qquad -  q^{\frac 12}  \sum_{e \in {\rm Edge}_{\gamma'}}   \left( {\CalW}_{{\bnu}, t(e), {\ib}''} {\CalK}_{{\bnu}+{\bro}, s(e), {\ib}''}^{*} +
{\CalW}_{{\bnu}, s(e), {\ib}''} {\CalK}_{{\bnu}+{\bro}, t(e), {\ib}''}^{*} \right) \right\} \\
\end{multline}

\subsec{The main fact}

For all $\Gammadi$, let us denote by
\beq
x_{A} = \frac{1}{\sum_{{\bom} \in {\Gammadi}^{\vee}} n_{A, {\bom}}} \sum_{{\bom}\in {\Gammadi}^{\vee}}  \sum_{{\alpha} \in [n_{A, {\bom}}]} {\ac}_{A, {\bom}}
\label{eq:xaom}
\eeq
The partition function of the orbifold gauge origami, defined by \eqref{eq:abelorb} in the {\bf abelian} case, by 
\beq
{\CalZ}_{\rm cross}^{\Gammadi} ({\ba}, {\bb}; {\ve}_{1}, {\ve}_{2}; {\bqe}) = \sum_{{{\bla}_{{\rm ab}\times {\rm ale}}}} {\CalZ}^{\rm pert}_{\Gammadi, \rm cross} z_{{\bla}_{{\rm ab}\times {\rm ale}}}\eeq
 in the  {\bf abelian} $\times$ {\bf ALE} case, 
\beq
{\CalZ}_{\rm cross}^{\Gammadi} ({\ba}, {\bb}; {\ve}; {\bqe}) = \sum_{{\bla}_{{\rm ale}\times {\rm ale}}} {\CalZ}^{\rm pert}_{\Gammadi, \rm cross} z_{{\bla}_{{\rm ale}\times {\rm ale}}}
\eeq
in the {\bf ALE}$\times${\bf ALE} case, 
has no singularities in the $x_{A}$ variables, with ${\tilde\ac}_{A, {\bom}} = {\ac}_{A, {\bom}} - x_{A}$
fixed. Again, this follows from the compactness theorem proven in \cite{Nekrasov:2016qym}. 

\secc{Conclusions\ and\ outlook}
 
The partition function of the gauge origami model,  can be viewed as
the expectation value in the ${\CalN}=2^{*}$ $U(n_{A})$ theory on ${\BC}^{2}_{A}$ of an operator. In the crossed case,  $N_{A}N_{B} = 0$, $A \cap B \neq {\emptyset}$, this operator is the qq-character of the ${\hat A}_{0}$-type \cite{Nekrasov:2015wsu}. In the orbifolded crossed case this operator is the qq-character of the ${\hat{\mathfrak g}}_{\gamma}$-type.  
 The orbifold partition functions in the {\bf abelian} case describe the ${\hat A}$-type quiver gauge theories on the $A$-type ALE spaces in the presence of various surface defects invariant under the rotational symmetries of the maximal $\Omega$-deformation. In the {\bf abelian}$\times${\bf ALE} case these partition functions describe either the qq-characters of the ${\hat D}$ or ${\hat E}$-type quiver gauge theories, possibly with the surface defects, or
 the ${\hat A}$-type quiver gauge theory on the $D$ or $E$-type ALE space, possibly with 
 a novel type of surface defect (which collapses to a point-like defect in the orbifold limit of the ALE space), and a qq-character. Finally, in the {\bf ALE}$\times${\bf ALE} case we are dealing with the  
 ${\hat D}$ or ${\hat E}$-type quiver gauge theories, on the $D$ or $E$-type ALE space, with the qq-characters and novel surface defects. 
 
 The physics of these defects will be discussed in the companion paper \cite{NeIV}. 
 
The regularity of these expectation values will be used in the forthcoming publications \cite{NeV, SX} to derive the KZ and BPZ  equations \cite{Belavin:1984vu, Knizhnik:1984} on the partition functions of supersymmetric gauge theories 
 with and without surface operators.

\end{document}